\documentclass[12pt]{article}
\usepackage[dvips]{graphicx}
\usepackage{amsfonts}
\usepackage{amssymb}

\title{Applications of density matrices \\ in a trapped Bose gas}

\author{K.~Ch.~Chatzisavvas\footnote{\texttt{e-mail:\, kchatz\,@\,auth.gr}}\,,
        S.~E.~Massen\footnote{\texttt{e-mail:\, massen\,@\,auth.gr}}\,,
        Ch.~C.~Moustakidis\footnote{\texttt{e-mail:\, moustaki\,@\,auth.gr}}\,,
        C.~P.~Panos\footnote{\texttt{e-mail:\,
        chpanos\,@\,auth.gr}}
\\
 {\it  Department of Theoretical Physics,}\\
        {\it Aristotle University of Thessaloniki,}\\
                {\it  54124 Thessaloniki, Greece}
 }

\date{January, 2006}

\begin{document}

\maketitle

\begin{abstract}
An overview of the Bose-Einstein condensation of correlated atoms
in a trap is presented by examining the effect of interparticle
correlations to one- and two-body properties of the above systems
at zero temperature in the framework of the lowest order cluster
expansion. Analytical expressions for the one- and two-body
properties of the Bose gas are derived using Jastrow-type
correlation function. In addition numerical calculations of the
natural orbitals and natural occupation numbers are also carried
out. Special effort is devoted for the calculation of various
quantum information properties including Shannon entropy, Onicescu
informational energy, Kullback-Leibler relative entropy and the
recently proposed Jensen-Shannon divergence entropy. The above
quantities are calculated for the trapped Bose gases by comparing
the correlated and uncorrelated cases as a function of the
strength of the short-range correlations. The Gross-Piatevskii
equation is solved giving the density distributions in position
and momentum space, which are employed to calculate quantum
information properties of the Bose gas.
\end{abstract}


\section{Introduction}\label{sec:sec1}

The first theoretical prediction of the famous phenomenon known as
Bose-Einstein condensation (BEC) was made in 1924 and 1925 by Bose
\cite{Bose24} and Einstein \cite{Einstein 24-25}, respectively. In
a system of particles obeying Bose-Einstein statistics where the
total number of particles is conserved, there should be a
temperature below which a finite fraction of all the particles
condense into the same one-particle state
\cite{Griffin95,Dalfavo99,Leggett01,Parkins98,Fetter98,Courteille01}.
Seventy years later, in a remarkable experiment, Anderson \emph{et
al.} \cite{Anderson95} have cooled magnetically trapped $^{87}$Rb
gas to nanokelvin temperatures, and observed the BEC. This
discovery has generated a huge amount of theoretical
investigations
\cite{Baym96,Dalfavo96,Esry97,Fabrocini99,Fabrocini01,DuBois01,Minguzzi00,Naraschewski99,Pitaevski99}.

The main feature of the trapped alkali-metal and atomic hydrogen
systems (which obey the Bose-Einstein statistics) is that they are
dilute. The crucial parameter defining the condition of diluteness
is the gas parameter $\chi=na^3$, where $n$ is the density of the
system and $a$ is the s-wave scattering length \cite{Fabrocini01}.
There are two ways to bring $\chi$ outside the regime of validity
of the mean field description. The first one is to increase the
density, while the second one to change the effective size of the
atoms.

The diluteness of the gas is ensured when the effective atomic
size is small compared both to the trap size and to the
interatomic distance.  However, the effects of inter-particle
interactions are of fundamental importance in the study of the BEC
dilute-gas where the physics should be dominated by two-body
collisions described in terms of the s-wave scattering length $a$.
In the case of positive $a$, it is equivalent to consider a very
dilute  (atomic) system of hard spheres, whose diameter coincides
with the scattering length itself \cite{Fabrocini99}. The natural
starting point for studying the behavior of those systems is the
theory of weakly interacting bosons which, for inhomogeneous
systems, takes the form of the Gross-Pitaevskii equation
\cite{Pitaevski61,Gross61}. This is a mean-field approach for the
order parameter associated with the condensate \cite{Dalfavo99}.

In the present work we study BEC in a phenomenological way where
the Bose gas is considered as a many-body system
\cite{Ketterle99}. In particular, we study the ground state of a
system of correlated bosonic atoms at zero temperature, trapped by
a harmonic oscillator potential (HO). The key quantities for this
effort is the one- and two-body density matrices \cite{Lowdin}. As
the mean-field approach (non-interacting atoms) fails to
incorporate the interparticle interactions which are necessary for
the description of the correlated Bose system, we introduce the
repulsive interactions among the atoms, through the Jastrow
correlation functions $f\left(|\textbf{r}_1-\textbf{r}_2| \right)$
\cite{Moustakidis02}.

We focus our efforts to the calculation of the one- and two-body
density and momentum distributions and the calculation of the
static structure factor
\cite{Moustakidis02,MoustaChatz05,Moustakidis04,Massen05}.
One-body density and momentum distributions are complementary
descriptions of the Bose gas and related directly with the
mean-square radius and mean kinetic energy of the trapped Bose gas
respectively. In addition the two-body density distribution is
related with the calculation of the static structure factor, a
quantity which gives information for the ground and excited states
of the gas. Special effort has been devoted to the derivation of
the natural orbital and natural occupation numbers through the
diagonalization of the one-body density matrix.

 In recent years information-theoretic methods play an increasing
role for the study of quantum mechanical systems. An example is
the application of the Maximum Entropy Principle \cite{Kapur89}
(MEP) to the calculation of the wave function in a potential
\cite{Canosa92} using as constraints expectation values of simple
observables and reconstructing a quantum wave function from a
limited set of expectation values. The idea behind MEP is to
choose the least biased result, compatible with the constraints of
the problem. Thus the MEP provides the least biased description
consistent with the available relevant information. This is done
by employing a suitably defined quantum entropy that measures the
lack of information associated with the distribution of a quantum
state over a given known basis.

Information entropy is important for the study of quantum
mechanical systems in two cases: first in the clarification of
fundamental concepts of quantum mechanics and second in the
synthesis of probability densities in position and momentum space
\cite{Garbaczewski05}.

In the present work special effort is devoted for the calculation
of various quantum information properties including Shannon
entropy, Onicescu energy, Kullback-Leibler relative entropy and
also Jensen-Shannon divergence. The information properties are
calculated for the interatomic correlations. In addition the
Gross-Piatevskii equation is solved giving the density and
momentum distribution which are employed to calculate the above
quantum information properties of the Bose gas
\cite{Massen05,Massen02}. The results are compared with those
taken in a phenomenological way in the framework of the Jastrow
correlations.

The plan of the paper is the following: In Sec. \ref{sec:sec2} the
general definitions related to the density matrices of a Bose
system are considered. Details of the lowest-order cluster
expansion, analytical expressions and numerical results are
reported in Sec. \ref{sec:sec3}. In Sec. \ref{sec:sec4} formulas
for the quantum information properties (both for the one-and
two-body density matrices) are reviewed and analytical results are
presented. Quantum information properties based on
Gross-Piatevskii equation are presented in Sec. \ref{sec:sec5}
while the summary of the work is given in Sec. \ref{sec:sec6}.

\section{Definition of Density Matrices}\label{sec:sec2}

Let $\Psi({\bf r}_1,{\bf r}_2,\ldots,{\bf r}_A)$ be the wave
function describing the trapped Bose gases. In the case where this
system is composed of non-interacting bosonic atoms at zero
temperature, all atoms occupy the same single-particle ground
state. The many body ground state wave function $\Psi_0({\bf
r}_1,{\bf r}_2,\ldots,{\bf r}_A)$ is then a product of $A$
identical single particle ground state wave functions. This ground
state wave function is therefore called the condensate wave
function or macroscopic wave function and has the form
\cite{Ketterle99}
\begin{equation}
\Psi_0({\bf r}_1,{\bf r}_2,\ldots,{\bf r}_A)= \psi_0({\bf
r}_1)\psi_0({\bf r}_2) \cdots \psi_0({\bf r}_A), \label{WF-1}
\end{equation}

\noindent where $\psi_0({\bf r})$ is the normalized to one
ground-state single-particle wave function describing a bosonic
atom. It is worth to indicate that Eq. (\ref{WF-1}) is valid even
when
 weak interactions are included. In this case the wave function
$\Psi_0({\bf r}_1,{\bf r}_2,\ldots,{\bf r}_A)$ is still, to a very
good approximation, a product of $A$ single particle wave
functions obtained now from the solution of a non-linear
Schr\"odinger equation, the well known Gross-Pitaevskii equation.
However, in the general case where interactions between atoms are
included, the ground state wave function $\Psi({\bf r}_1,{\bf
r}_2,\ldots,{\bf r}_A)$ is modified from the simple form of Eq.
(\ref{WF-1}). In that case a percentage of atoms is moving from
the condensate orbit $\psi_0$ to higher orbits.

In the present work we adopt the following normalization of the
wave function $\Psi({\bf r}_1,{\bf r}_2,\ldots,{\bf r}_A)$,

\begin{equation}
 \int \Psi^{*}({\bf r}_1,{\bf r}_2,\ldots,{\bf r}_A)
 \Psi({\bf r}_1,{\bf r}_2,\ldots,{\bf r}_A)
 d\textbf{r}_1 d\textbf{r}_2 \cdots d\textbf{r}_A=1,
\end{equation}

\noindent where the integration is carried out over the radius
vectors $\textbf{r}_1$, $\textbf{r}_2$, $\ldots$, $\textbf{r}_A$.

A quantity characterizing very important aspects of a Bose gas (as
well a variety of quantum many-body systems) is the one-body
density matrix defined as in \cite{Lowdin}

\begin{equation}
\rho({\bf r}_1,{\bf r}_1')=\int \Psi^{*}({\bf r}_1,{\bf
r}_2,\ldots,{\bf r}_A) \Psi({\bf r}_1',{\bf r}_2,\ldots,{\bf r}_A)
d{\bf r}_2 \cdots d{\bf r}_A. \label{OBDM-1}
\end{equation}

The one-body density matrix is connected to the position- and
momentum-space properties of the Bose gas and in addition it is
the quantity which gives the percentage of the condensate of the
system.

The two-body density matrix is a generalization of the one-body
density matrix and is defined as

\begin{equation}
\rho({\bf r}_1,{\bf r}_2;{\bf r}_1',{\bf r}_2')= \int
\Psi^{*}({\bf r}_1,{\bf r}_2,{\bf r}_3,\cdots,{\bf r}_A) \Psi({\bf
r}_1',{\bf r}_2',{\bf r}_3,\cdots,{\bf r}_A) d{\bf r}_3 \cdots
d{\bf r}_A. \label{TBDM-1}
\end{equation}
The above density matrices are related by the following equation
\begin{equation}
\rho({\bf r}_1,{\bf r}_1')=\int \rho({\bf r}_1,{\bf r}_2;{\bf
r}_1',{\bf r}_2)  d{\bf r}_2. \label{O-T-1}
\end{equation}

The two-body density matrix is related directly to the interatomic
interaction and its diagonal part provides the two-body density
distribution $\rho(\textbf{r}_1,\textbf{r}_2)$ (expresses the
joint probability of finding two atoms at the positions ${\bf
r}_1$ and ${\bf r}_2$, respectively), a key quantity of the
present work

\begin{equation}
\rho({\bf r}_1,{\bf r}_2)= \rho({\bf r}_1,{\bf r}_2;{\bf
r}_1',{\bf r}_2')\mid_{{\bf r}_1'={\bf r}_1, {\bf r}_2'={\bf
r}_2}. \label{TBDD-1}
\end{equation}

On the other hand the diagonal part of the one-body density matrix
is just the density distribution of the Bose gas and expresses the
probability of finding an atom at position $\textbf{r}_1$

\begin{equation}
\rho({\bf r}_1)=\rho({\bf r}_1,{\bf r}_1')|_{{\bf r}_1={\bf
r}_1'}. \label{DD-1}
\end{equation}

The quantities $\rho(\textbf{r}_1)$ and
$\rho(\textbf{r}_1,\textbf{r}_2)$ are also related by the
following integral

\begin{equation}
\rho({\bf r}_1)=\int \rho({\bf r}_1,{\bf r}_2)
 d{\bf r}_2. \label{DD-2}
\end{equation}

Very interesting is also the description of the Bose gas in
momentum-space via the quantities of the one- and two-body
momentum distributions. The two-body momentum distribution
$n(\textbf{k}_1,\textbf{k}_2)$ expresses the joint probability of
finding two atoms with momentum $\textbf{k}_1$ and $\textbf{k}_2$
respectively and is given by a particular Fourier transform of the
corresponding two-body density matrix $\rho({\bf r}_1,{\bf
r}_2;{\bf r}_1',{\bf r}_2')$

\begin{equation}
n({\bf k}_1,{\bf k}_2)=\frac{1}{(2\pi)^6} \int \rho({\bf r}_1,{\bf
r}_2;{\bf r}_1',{\bf r}_2') \exp[i{\bf k}_1({\bf r}_1-{\bf r}_1')]
\exp[i{\bf k}_2({\bf r}_2-{\bf r}_2')] d {\bf r}_1 d {\bf r}_1' d
{\bf r}_2 d {\bf r}_2'. \label{TBMD-1}
\end{equation}

The one-body momentum distribution (or simply momentum
distribution) $n(\textbf{k})$, expresses the probability of
finding an atom with momentum $\textbf{k}$, and it is given by a
particular Fourier transform of the one-body density matrix
$\rho({\bf r}_1,{\bf r}_1')$

\begin{equation}
n({\bf k})=\frac{1}{(2\pi)^3} \int \rho({\bf r}_1,{\bf r}_1') \exp
\left[i{\bf k}({\bf r}_1-{\bf r}_1')\right]
 d  {\bf r}_1   d  {\bf r}_1'. \label{mom}
\end{equation}

It can be shown easily that in the case where the Bose gas is
described by the wave function of Eq. (\ref{WF-1}) the two-body
density matrix is given by
\begin{equation}
\rho_{0}({\bf r}_1,{\bf r}_2;{\bf r}_1',{\bf r}_2')= \rho_{0}({\bf
r}_1,{\bf r}_1')\rho_{0}({\bf r}_2,{\bf r}_2'), \label{TBDM-BG}
\end{equation}
where
\begin{equation}
\rho_{0}({\bf r}_1,{\bf r}_1')=\psi_{0}^{*}({\bf r}_1)
\psi_{0}({\bf r}_1'). \label{eq:eq12new}
\end{equation}

From Eqs. (\ref{TBMD-1}), (\ref{mom}) and (\ref{TBDM-BG}) we get
\begin{eqnarray}
n_0(\textbf{k}_1,\textbf{k}_2) &=& \displaystyle{
\frac{1}{(2\pi)^3}} \int \rho_0(\textbf{r}_1,\textbf{r}_1')\,
\textrm{exp}\left[ i \textbf{k}_1 (\textbf{r}_1-\textbf{r}_1')
\right]\,d\textbf{r}_1
d\textbf{r}_1' \nonumber \\
& & \times \displaystyle{ \frac{1}{(2\pi)^3}} \int
\rho_0(\textbf{r}_2,\textbf{r}_2')\, \textrm{exp}\left[ i
\textbf{k}_2 (\textbf{r}_2-\textbf{r}_2') \right]\,d\textbf{r}_2
d\textbf{r}_2' \\
&=& n_0(\textbf{k}_1)n_0(\textbf{k}_2). \nonumber
\end{eqnarray}

Finally form Eqs. (\ref{mom}) and (\ref{eq:eq12new}) we get

\begin{eqnarray}
n_0(\textbf{k})&=& \displaystyle{ \frac{1}{(2\pi)^{3/2}}} \int
\psi_{0}^{*}({\bf r}_1) \textrm{exp}\left[ i \textbf{k}
\textbf{r}_1 \right] d\textbf{r}_1 \nonumber \\
& & \times \displaystyle{ \frac{1}{(2\pi)^{3/2}}} \int
\psi_{0}^{*}({\bf r}_1') \textrm{exp}\left[ i \textbf{k}
\textbf{r}_1' \right] d\textbf{r}_1' \label{eq:eq14new} \\
&=& \tilde{\psi}_{0}^{*}({\bf k})\tilde{\psi}_{0}({\bf k}).
\nonumber
\end{eqnarray}

From Eq. (\ref{eq:eq14new}) it is obvious that
$\tilde{\psi}_{0}^{*}({\bf k})$ is the particular Fourier
transform of the single particle wave function $\psi_{0}({\bf
r})$.

\subsection{Static Structure Factor}

Spectroscopic studies have been used to assemble a complete
understanding of the structure of atoms and simple molecules
\cite{Stamper-Kurn}. The static structure factor $S(k)$ is a
fundamental quantity, connected with the atomic structure, and is
the Fourier transform of the radial distribution function $g(r)$.
$S(k)$ gives the magnitude of the density fluctuation in the
system (atomic, molecular, electronic or nuclear) at wavelength
$2\pi/k$, where $k$ is the momentum transfer. In recent papers,
the Bragg spectroscopic method was used  to measure $S(k)$ either
in the phonon regime \cite{Stamper-Kurn} or/and in the
single-particle regime \cite{Steinhauer}.

The static structure factor in a finite system is defined as
\cite{Zambelli}

\begin{equation}
S({\bf k})=1+ \frac{1}{N} \int e^{i{\bf k}({\bf r}_1-{\bf r}_2)}
\left[ \rho({\bf r}_1,{\bf r}_2) - \rho({\bf r}_1) \rho({\bf r}_2
) \right] d  {\bf r}_1   d  {\bf r}_2. \label{eq:eq18}
\end{equation}

In the most general case the two-body density distribution
$\rho(\textbf{r}_1,\textbf{r}_2)$ and the one-body density
distribution $\rho(\textbf{r})$ are connected via the following
relation

\begin{equation}
 \rho(\textbf{r}_1,\textbf{r}_2)=C\rho(\textbf{r}_1)\rho(\textbf{r}_2)f^2(r_{12})
 =C\rho(\textbf{r}_1)\rho(\textbf{r}_2)g(r_{12}), \label{eq:eq19}
\end{equation}
where $g(r_{12})$ is the radial distribution function and $C$ is
the normalization factor which ensures that

\begin{equation}
  \int \rho(\textbf{r}_1,\textbf{r}_2) d\textbf{r}_1
  d\textbf{r}_2=N(N-1). \label{eq:eq20}
\end{equation}

We also consider that
\begin{equation}
  \int \rho(\textbf{r}_1) d\textbf{r}_1=N, \label{eq:eq21}
\end{equation}
where $N$ is the number of the atoms of the Bose condensate.

In the uncorrelated case (non-interacting gas) the radial
distribution function is $g(r_{12})=1$ (absence of correlations),
and the two-body density distribution becomes

\begin{equation}
 \rho(\textbf{r}_1,\textbf{r}_2)=
 \frac{N-1}{N}\rho(\textbf{r}_1)\rho(\textbf{r}_2).
\end{equation}

Using Eq. (\ref{eq:eq19}), Eq. (\ref{eq:eq18}) is written as
\begin{equation}
S({\bf k})=1+ \frac{1}{N} \int \textrm{exp} \left[i{\bf k}({\bf
r}_1-{\bf r}_2)\right] \rho({\bf r}_1) \rho({\bf r}_2 ) [C
g(r_{12})-1] d {\bf r}_1   d {\bf r}_2. \label{str-fin2}
\end{equation}

Conditions (\ref{eq:eq20}) and (\ref{eq:eq21}) ensure that
$S(0)=0$.

The integration in Eq. (\ref{str-fin2}) can be performed if the
function $g(r)$ is known. $g(r)$ must obey  the rules $g(r=0)=0$
and $\displaystyle{\lim_{r \rightarrow \infty} g(r) \rightarrow
1}$. The first rule introduces the repulsive correlations between
the atoms and the second the absence of such correlations in long
distances. In general the form of $S(k)$ is affected appreciably
from the form of $g(r)$. More specifically, the long range
behavior of $g(r)$ affects $S(k)$ for small values of $k$ while
its short range behavior affects $S(k)$ for large values of $k$ as
a direct consequence of the Fourier transform theory.

\subsection{Natural Orbitals and Natural Occupation Numbers}

In the case of the inclusion of the inter-particle interactions
between the atoms, which give rise to the depletion of the
condensate, the one-body density matrix is written
\cite{Stringari01}
\begin{equation}
\rho({\bf r}_1,{\bf r}_1')= n_0 \psi_{0}^{*}({\bf r}_1)
\psi_{0}({\bf r}_1')+ \sum_{i\neq 0}n_i \psi_{i}^{*}({\bf r}_1)
\psi_{i}({\bf r}_1'),
\end{equation}
where $\displaystyle{\sum_{i}n_i=1}$. The sum
$\displaystyle{\sum_{i\neq 0}n_i \psi_{i}^{*}({\bf r}_1)
\psi_{i}({\bf r}_1')}$ is the contribution arising from the atoms
out of the condensate. The eigenfunctions $\psi_i({\bf r}),$ which
are called natural orbitals (NO's), and the eigenvalues $n_i$,
called natural occupation numbers (NON's), are obtained by
diagonalizing the one-body density matrix through the eigenvalue
equation
\begin{equation}
\int \rho({\bf r}_1,{\bf r}_1') \psi_i({\bf r}_1')  d {\bf r}_1'
=n_i\psi_i({\bf r}_1). \label{diag-rho}
\end{equation}
The condition, generally adopted, for the existence of
condensation is that there should be one eigenvalue $n_i$ which is
of the order of the number of the particles in the trap.

The NO's $\psi_{i}({\bf r}_1)$ and the NON's $n_i$ are obtained by
diagonalizing the one-body density matrix through the eigenvalue
equation (\ref{diag-rho}) by expanding first the one-body density
matrix in a series of Legendre polynomials $P_l(x)$
\begin{equation}
\rho({\bf r},{\bf r}')=\rho(r, r',\cos\omega_{rr'})=
\sum_{l=0}^{\infty}\rho_l(r,r') P_l(\cos\omega_{rr'}),
\label{ch4-NO4}
\end{equation}
where $\rho_l(r,r')$ are the  coefficients of the expansion
\begin{equation}
\rho_l(r,r')=\frac{2l+1}{2}\int_{-1}^{1} \rho(r,
r',\cos\omega_{rr'}) \ P_l(\cos\omega_{rr'}) \ {\rm d}
(\cos\omega_{rr'}). \label{ch4-NO5}
\end{equation}
From the Eqs. (\ref{diag-rho}), (\ref{ch4-NO4}) and
(\ref{ch4-NO5}) the eigenvalue equation is written
\begin{equation}
4\pi \int_{0}^{\infty}\rho_{l}(r,r')\varphi_{nl}^{NO}(r')
{r'}^2{\rm d} r'=n_{nl}^{NO}\varphi_{nl}^{NO}(r), \label{NO-3}
\end{equation}
where $\varphi_{nl}^{NO}(r)$ is the radial part of
 $\psi_{i}({\bf r})$
($\psi_{i}({\bf r})=\varphi_{nl}^{NO}(r)Y_{lm}(\Omega_r) $).

\section{Jastrow type Correlated Properties of a Trapped Bose
Gas}\label{sec:sec3}

\subsection{Correlated Density Matrices}\label{sub:sub3-1}

A dilute trapped Bose gas can be studied using the lowest-order
approximation \cite{Fabrocini99}. In this approximation the
two-body density matrix has the form
\cite{Moustakidis02,MoustaChatz05}
\begin{equation}
\rho({\bf r}_1,{\bf r}_2;{\bf r}_1',{\bf r}_2')= N_0 \rho_0({\bf
r}_1,{\bf r}_1' ) \rho_0({\bf r}_2,{\bf r}_2') f(|{\bf r}_1' -
{\bf r}_2'|) f(|{\bf r}_1-{\bf r}_2|), \label{TBDM-1}
\end{equation}
where $f(|{\bf r}_1-{\bf r}_2|)$ is the Jastrow correlation
function, which depends on the inter-particle distance and $N_0$
is the normalization factor which ensures that
\[\int \rho({\bf
r}_1,{\bf r}_2;{\bf r}_1',{\bf
r}_2')|_{(\textbf{r}_1=\textbf{r}_1', \textbf{r}_2=\textbf{r}_2')}
\, d\textbf{r}_1 d\textbf{r}_2=1.
\]

The diagonal part of $\rho({\bf r}_1,{\bf r}_2;{\bf r}_1',{\bf
r}_2')$ that is the two body density distribution takes the form
\begin{equation}
\rho({\bf r}_1,{\bf r}_2)= N_0 \rho_0({\bf r}_1)\rho_0({\bf r}_2)
f^2(r_{12}), \label{TBDD-12}
\end{equation}
while the one-body density matrix is given by the integral
\begin{equation}
\rho({\bf r}_1,{\bf r}_1')= N_0 \rho_0({\bf r}_1,{\bf r}_1') \int
\rho_0({\bf r}_2,{\bf r}_2) f(|{\bf r}_1' - {\bf r}_2|)f(|{\bf
r}_1 - {\bf r}_2|) d\textbf{r}_2. \label{OBDM-1}
\end{equation}

The density distribution, which is the diagonal part of $\rho({\bf
r}_1,{\bf r}_1')$, can also be obtained from the integral
\begin{eqnarray}
\rho({\bf r})= \int \rho({\bf r},{\bf r}_2) d{\bf r}_2 =N_0
\rho_0({\bf r}) \int \rho_0({\bf r}_2) f^2(|{\bf r}-{\bf r}_2|)  d
{\bf r}_2. \label{cor-dd1}
\end{eqnarray}

In the present work we consider that the atoms are confined in an
isotropic HO well where the normalized to $1$ ground state single
particle wave function $\psi_0(r)$ has the form of a Gaussian
given by the formula
\[
  \psi_0(r)=\left( \frac{1}{\pi b^2} \right)^{3/4}
  \textrm{exp}\left[-\frac{r^2}{2 b^2}\right], \quad
  \mbox{\textrm{where the width}}\,\, b=\left( \frac{\hbar}{m\omega}
  \right)^{1/2},
\]
while the density distribution has the form  $\rho_0({\bf r})=|
\psi_0({\bf r}) |^2$. The correlation function $f(r_{12})$ is
taken to be of the form
\begin{equation}
f(r)=1-\exp\left[-\frac{y r^{2}}{b^2}\right], \label{case-1}
\end{equation}
where $r=|{\bf r}_1-{\bf r}_2|$. The correlation function  $f(r)$
goes to $1$ for large values of $r$ and goes to $0$ for $r
\rightarrow 0$. It is obvious that the effect of the correlations
introduced by the function $f(r)$, becomes large when the
correlation parameter $y$ becomes small and vice versa.

The above defined correlation function was used in
\cite{Moustakidis02,MoustaChatz05} to find analytical expressions
of the one-body density matrices in position and momentum spaces
and static structure factor, while the NO's and NON's are
calculated numerically employing Eq. (\ref{NO-3}).

The analytical expression of the two-body density matrix obtained
from Eq. (\ref{TBDM-1}) has the form

\begin{eqnarray}
 \rho(\textbf{r}_1,\textbf{r}_2,\textbf{r}'_1,\textbf{r}'_2)= &
 \displaystyle{\frac{N_0}{\pi^3 b^6}} \, \Big(
 O_1(\textbf{r}_1,\textbf{r}_2,\textbf{r}'_1,\textbf{r}'_2)-
 O_{21}(\textbf{r}_1,\textbf{r}_2,\textbf{r}'_1,\textbf{r}'_2)- \nonumber \\
 &  -O_{22}(\textbf{r}_1,\textbf{r}_2,\textbf{r}'_1,\textbf{r}'_2)+
 O_{23}(\textbf{r}_1,\textbf{r}_2,\textbf{r}'_1,\textbf{r}'_2)
 \Big),
\end{eqnarray}
where $N_0$ is the normalization factor of the form
\cite{MoustaChatz05}
\begin{equation}
N_0=\left[1-\frac{2}{(1+2y)^{3/2}}+\frac{1}{(1+4y)^{3/2}}\right]^{-1}
\label{norm-1b}
\end{equation}
and
\[
 \begin{array}{lcl}
 O_1(\textbf{r}_1,\textbf{r}_2,\textbf{r}'_1,\textbf{r}'_2)&=&
 \textrm{exp}\left[\displaystyle{-\frac{r_{1b}^2+r_{1b}^{'2}+r_{2b}^2+r_{2b}^{'2}}{2}}\right], \\
 O_{21}(\textbf{r}_1,\textbf{r}_2,\textbf{r}'_1,\textbf{r}'_2)&=&
 \textrm{exp}\left[\displaystyle{-\frac{r_{1b}^2+r_{2b}^2}{2}}\right]\,
 \textrm{exp}\left[\displaystyle{-\frac{(1+2y)(r^{'2}_{1b}+r^{'2}_{2b})}{2}}\right]\,
 \textrm{exp}\left[2y \textbf{r}'_{1b} \textbf{r}'_{2b}\right], \\
 O_{22}(\textbf{r}_1,\textbf{r}_2,\textbf{r}'_1,\textbf{r}'_2)&=&
 \textrm{exp}\left[\displaystyle{-\frac{r^{'2}_{1b}+r^{'2}_{2b}}{2}}\right]\,
 \textrm{exp}\left[\displaystyle{-\frac{(1+2y)(r_{1b}^2+r_{2b}^2)}{2}}\right]\,
 \textrm{exp}\left[2y \textbf{r}_{1b} \textbf{r}_{2b}\right], \\
 O_{23}(\textbf{r}_1,\textbf{r}_2,\textbf{r}'_1,\textbf{r}'_2)&=&
 \textrm{exp}\left[\displaystyle{-\frac{(1+2y)(r_{1b}^2+r_{1b}^{'2}+r_{2b}^2+r_{2b}^{'2})}{2}}\right]\,
 \textrm{exp}\left[2y(\textbf{r}_{1b}\textbf{r}_{2b}+\textbf{r}'_{1b}\textbf{r}'_{2b})\right].
\end{array}
\]
where $r_b=r/b$.

The two-body density distribution  in accordance with Eq.
(\ref{TBDD-1}) is given by

\begin{equation}
 \rho(\textbf{r}_1,\textbf{r}_2)=\frac{N_0}{\pi^3 b^6}
 \,\textrm{exp}[-r_{1b}^2]\,\textrm{exp}[-r_{2b}^2]\,
 \Big(1-\textrm{exp}[-y({\bf r}_{1b}-{\bf r}_{2b})^2]\Big)^2.
\end{equation}

The analytical expressions of the one-body density matrix obtained
from Eq. (\ref{OBDM-1}) has the form \cite{Moustakidis02}
\begin{equation}
\rho({\bf r},{\bf r}')= \frac{N_0}{\pi^{3/2}b^{3}}\,\left[O_1({\bf
r},{\bf r}')-O_{21}({\bf r},{\bf r}')- O_{22}({\bf r},{\bf
r}')+O_{23}({\bf r},{\bf r}')\right], \label{cluster-11}
\end{equation}

\noindent where the one- and the two-body terms of the expansion
in the low order approximation have the forms
\begin{eqnarray}
 O_1({\bf r},{\bf r}')&=&
\exp\left[-\frac{r_b^2+{r_b'}^2}{2}\right], \\
& & \nonumber\\
O_{21}({\bf r},{\bf r}')&=& \frac{1}{(1+y)^{3/2}}
\exp\left[-\frac{1+3y}{1+y}\frac{r_b^2}{2}-\frac{{r_b'}^2}{2}\right],  \\
& & \nonumber\\
O_{22}({\bf r},{\bf r}')&=&O_{21}({\bf r}',{\bf r}), \\
& & \nonumber\\
O_{23}({\bf r},{\bf r}')&=& \frac{1}{(1+2y)^{3/2}}
\exp\left[-(1+2y)\frac{r_b^2+{r_b'}^2}{2}\right]  \nonumber\\
&& \times \exp\left[\frac{y^2}{1+2y}({\bf r}_b+{\bf
r}_b')^2\right], \label{case1-o1r}
\end{eqnarray}

The analytical expression of the density distribution can be found
from Eq. (\ref{cluster-11}), putting ${\bf r'}={\bf r}$
\begin{eqnarray}
\rho(r)&=& \frac{N_0}{\pi^{3/2} b^3} \left(
\exp\left[-r_b^2\right] - \frac{2}{(1+y)^{3/2}}
\exp\left[-\frac{1+2y}{1+y}r_b^2\right]  \right.
\nonumber\\
&& \left. + \frac{1}{(1+2y)^{3/2}}
\exp\left[-\frac{1+4y}{1+2y}r_b^2\right]    \right).
\label{cluster-nr}
\end{eqnarray}

The two-body momentum distribution is calculated form the integral
of Eq. (\ref{TBMD-1}) and has the form

\begin{eqnarray}
 n(\textbf{k}_1,\textbf{k}_2)&=&\frac{b^6}{\pi^3} N_0 \,
 \textrm{exp}[-k_{1b}^2]\,\textrm{exp}[-k_{2b}^2]  \nonumber \\
 & & \times \left( 1-\frac{1}{(1+4y)^{3/2}}\,
 \textrm{exp}\left[-\frac{y}{1+4y}(\textbf{k}_{1b}-\textbf{k}_{2b})^2\right]
 \right)^2,
\end{eqnarray}

\noindent while the momentum distribution can be found
analytically using Eq. (\ref{mom}) and has the form
\begin{eqnarray}
n(k)&=& \frac{N_0 b^3}{\pi^{3/2}} \, \left(
\exp\left[-k_b^2\right] - \frac{2}{(1+3y)^{3/2}}
\exp\left[-\frac{1+2y}{1+3y}k_b^2\right]
\nonumber \right.\\
&& \left. + \frac{1}{(1+2y)^{3/2}(1+4y)^{3/2}}
\exp\left[-\frac{1}{1+2y}k_b^2\right]    \right),
\label{cluster-nk}
\end{eqnarray}
where $k_b =kb$.

The above analytical expressions of $\rho(r)$ and $n(k)$ have been
used to find the analytical expressions of the mean square radius
and kinetic energy of the trapped gas. The expressions we found,
for $\langle r^2 \rangle$ and $\langle T \rangle$, are
\begin{equation}
\langle r^2 \rangle=N_0 b^2 \left[ \frac{3}{2} -
3\frac{1+y}{(1+2y)^{5/2}} +\frac{3}{2}\frac{1+2y}{(1+4y)^{5/2}}
\right] \label{eq:rad-1}
\end{equation}
and
\begin{equation}
\langle T \rangle= N_0 \hbar \omega  \left[ \frac{3}{4} -
\frac{3}{2}\frac{1+3y}{(1+2y)^{5/2}} +
\frac{3}{4}\frac{1+2y}{(1+4y)^{3/2}}\right]. \label{eq:kinetic-1}
\end{equation}

These expressions, which for a given HO trap are functions of the
correlation parameter $y$, could be used to find the value of $y$
from Eq. (\ref{eq:rad-1}), if the rms radius of the trapped atoms
is known and then to define  $\langle T \rangle$ from Eq.
(\ref{eq:kinetic-1}) and vice versa. For very large values of $y$
Eqs. (\ref{eq:rad-1}) and (\ref{eq:kinetic-1}) give the HO
expressions of $\langle r^2 \rangle$ and $\langle T \rangle$, i.e.
$\langle r^2 \rangle = \frac{3}{2} b^2$ and $\langle T \rangle =
\frac{3}{4}\hbar\omega$, respectively.

\begin{figure}[h]
\centering
\includegraphics[height=5.0cm,width=4.0cm]{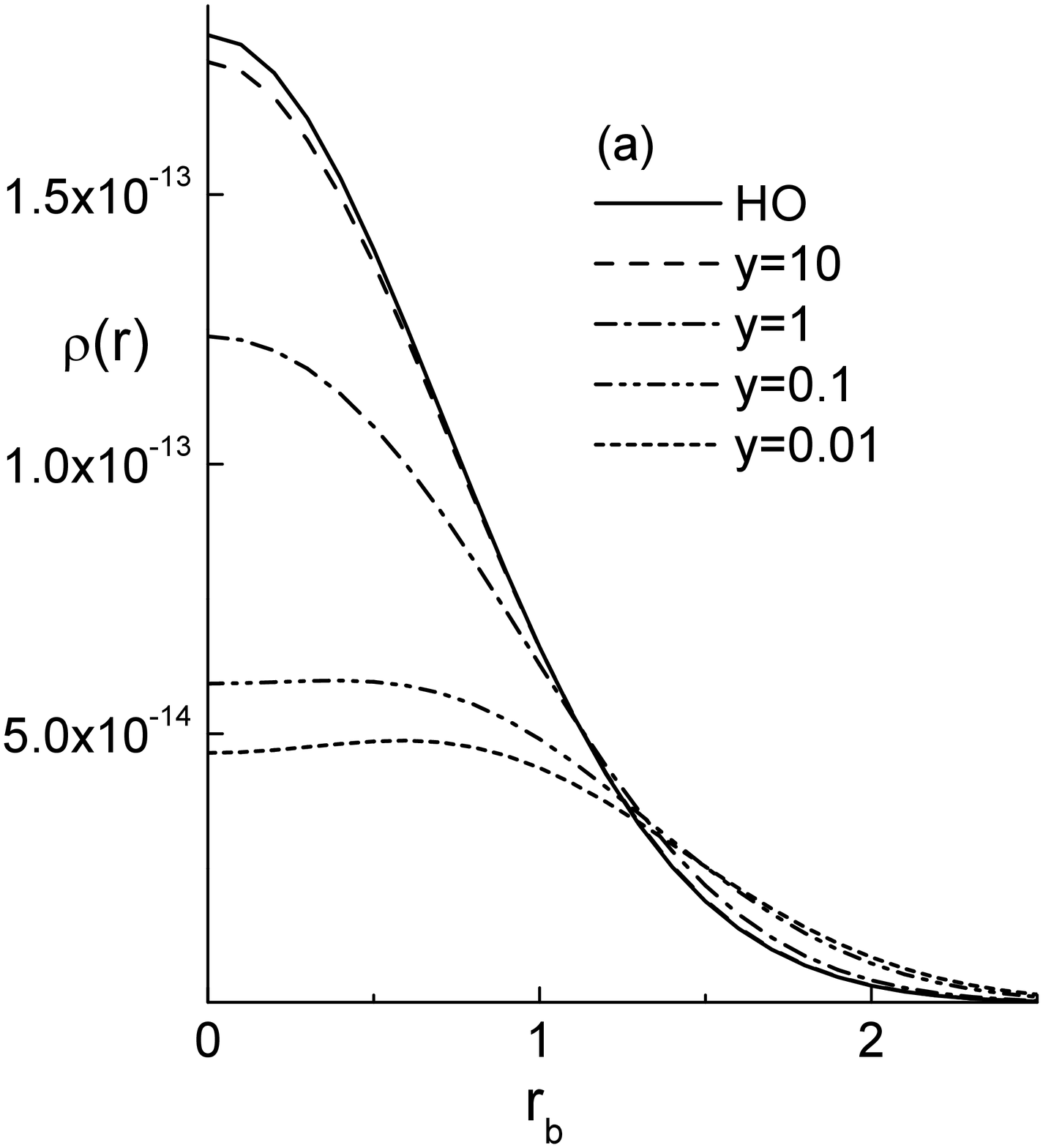}
\hspace{0.5cm}
\includegraphics[height=5.0cm,width=4.cm]{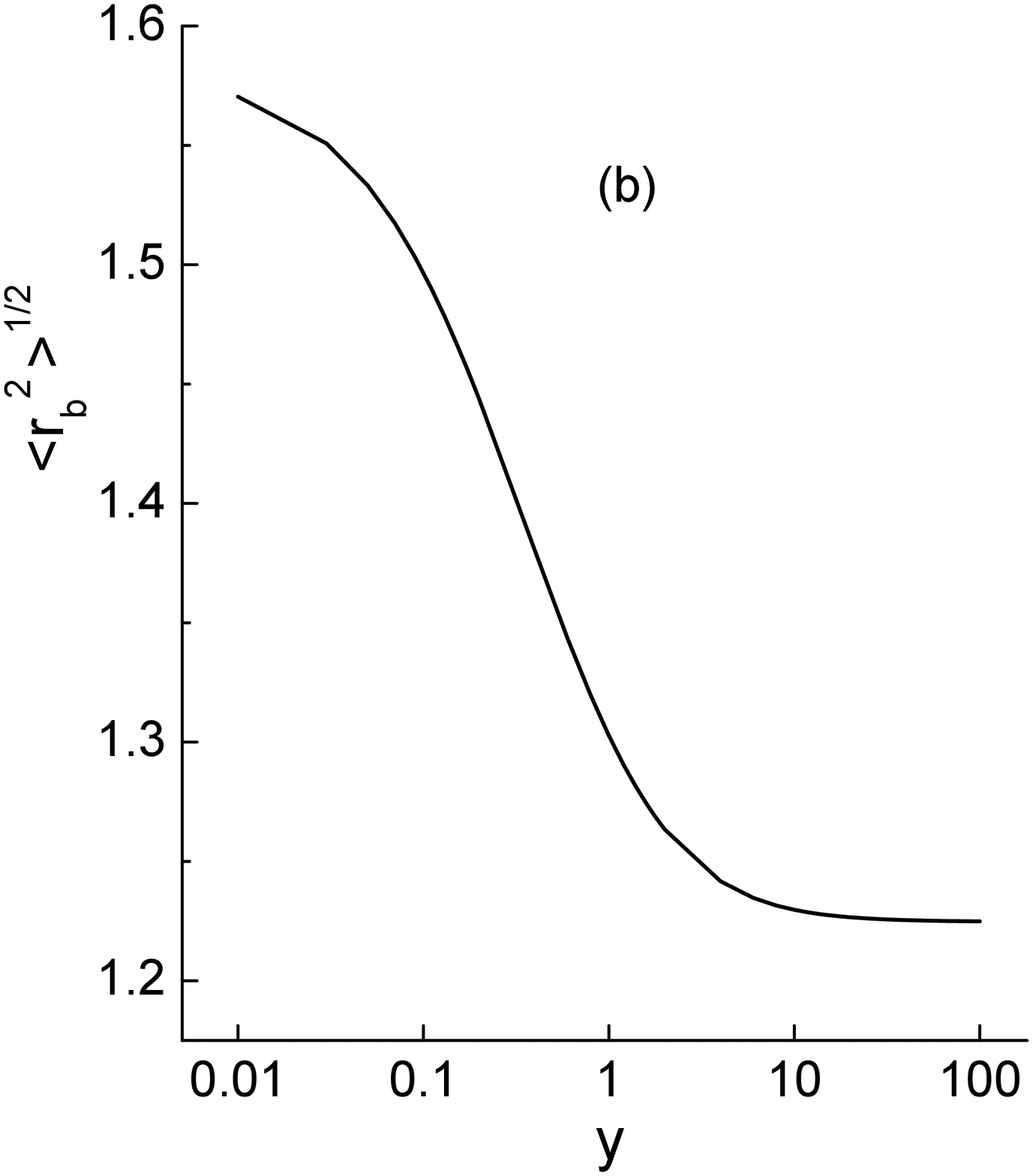} \caption{(a) The
density distribution $\rho(r)$ versus $r_b=r/b$ ($b=10^4$  \AA)
 for various
values of the parameter $y$. The normalization is $\int \rho({\bf
r})  d {\bf r}=1$. (b) The rms radius, $\sqrt{\langle r_b ^{2}
\rangle}$, versus the parameter $y$.}\label{fig:fig1}
\end{figure}

The calculation of the density distribution of a trapped Bose gas,
confined in an isotropic HO potential with length $b=10^4$ \AA,
has been carried out on the basis of Eq. (\ref{cluster-nr})
\cite{Moustakidis02}. The dependence of the density distribution
on the parameter $y$, including also the uncorrelated case
($y=\infty$), has been plotted in Fig. \ref{fig:fig1}(a). It is
seen that, the large values of $y$ ($y>10$) correspond to the
Gaussian distribution (HO case), while when $y$ becomes small
enough ($y<1$) the density distribution spreads out as in
Gross-Pitaevskii's theory \cite{Moustakidis02}. For $y>10$ the
effect of correlations is small, while for very large correlations
($y \lesssim 0.1$) the density distribution is modified entirely
compared to the Gaussian form originating from the HO trap.

The rms radius of the Bose gas has also been calculated
analytically from Eq. (\ref{eq:rad-1}) for various values of the
parameter $y$. From Fig. \ref{fig:fig1}(b), it is seen that
$\sqrt{\langle r_{b}^2\rangle}$ ($r_b=r/b$) is a decreasing
function of the parameter $y$ and for $y > 10$ approaches the rms
radius of the uncorrelated case.

\subsection{Natural Orbitals and Natural Occupations
Numbers}\label{sub:3-1}

\begin{figure}[t]
\centering
\includegraphics[height=5.0cm,width=4.0cm]{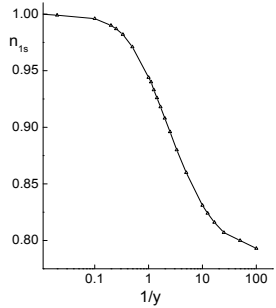}
\caption{The condensate fraction $n_{1s}$, at zero temperature,
for interacting atoms versus $1/y$.}\label{fig:fig2}
\end{figure}

The NO's and the NON's were calculated \cite{Moustakidis02}, by
diagonalizing the one-body density matrix through Eq.
(\ref{NO-3}). The NON $n_{1s}$, gives directly the condensation
fraction $n_0$ as a result of the repulsive interaction between
the atoms of the Bose gas at zero temperature. The NON's for the
$1s$, $1p$, $1d$ and $1f$ states are given in Table
\ref{tbl:table1}. It seems that, for strong correlations, a
fraction of atoms spread out into many states. The condensation
fraction $n_{1s}$, versus the parameter $\frac{1}{y}$ is plotted
in Fig. \ref{fig:fig2}. From that figure and from Table
\ref{tbl:table1} it is seen that the effect of the correlations on
$n_{1s}$ is small and all the atoms occupy the 1s ground state,
when $y>10$. The effect of the correlations is prominent when
$y<10$, while the decrease of the parameter $y$ (strong
correlations) induces a significant depletion of the condensated
atoms spreading them into many states.

\begin{table}[pt]
\centering
\begin{tabular}{@{}c c c c c c@{}}
\hline $y$ & $n_{1s}$ & $n_{1p}$ & $n_{1d}$  & $n_{1f}$ & Sum \\
\hline
100.00  & 0.99988  &   -       &    -     &  -         & 0.99988 \\
\hphantom{0}10.00   & 0.99634  & 0.00063   & 0.00042  &0.00042     & 0.99781  \\
\hphantom{00}5.00    & 0.99055  & 0.00273   & 0.00108  &0.00108     & 0.99544   \\
\hphantom{00}2.50  & 0.97771  & 0.00960   & 0.00186  &0.00186     & 0.99103   \\
\hphantom{00}1.00    & 0.94422  & 0.03462   & 0.00172  &0.00172     & 0.98228   \\
\hphantom{00}0.50  & 0.90815  & 0.06830   & 0.00082  &0.00082     & 0.97809   \\
\hphantom{00}0.10  & 0.83097  & 0.15185   & 0.00001  & 0.00001    & 0.98284  \\
\hphantom{00}0.01 & 0.79273  & 0.19414   &   -      & -          & 0.98687\\
\hline
\end{tabular}
\caption{The natural occupation numbers for various values of the
correlation parameter $y$ \cite{Moustakidis02}.}\label{tbl:table1}
\end{table}

\begin{figure}[t]
\centering
\includegraphics[height=5.0cm,width=3.5cm]{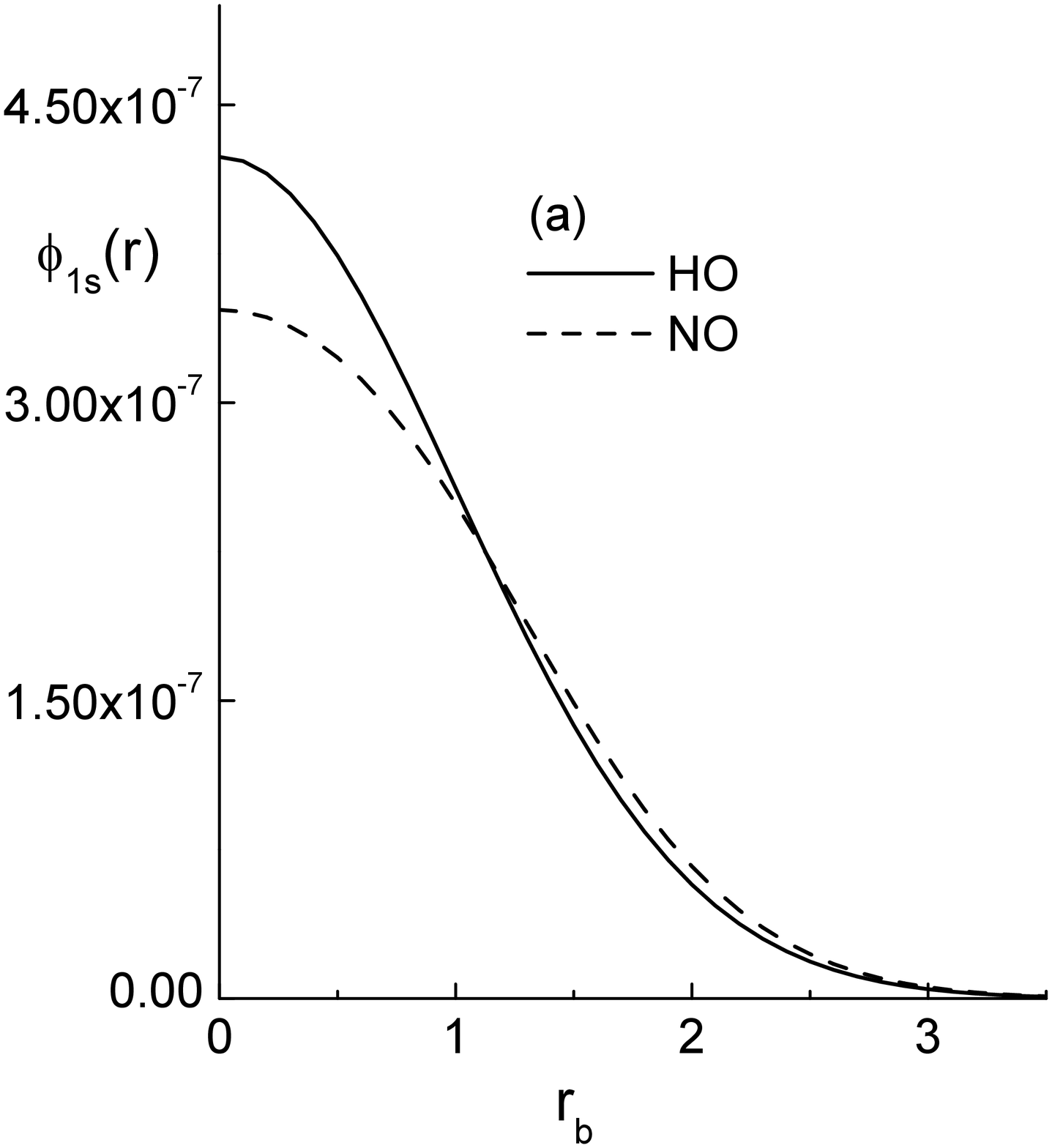}
\hspace{0.3cm}
\includegraphics[height=5.0cm,width=3.5cm]{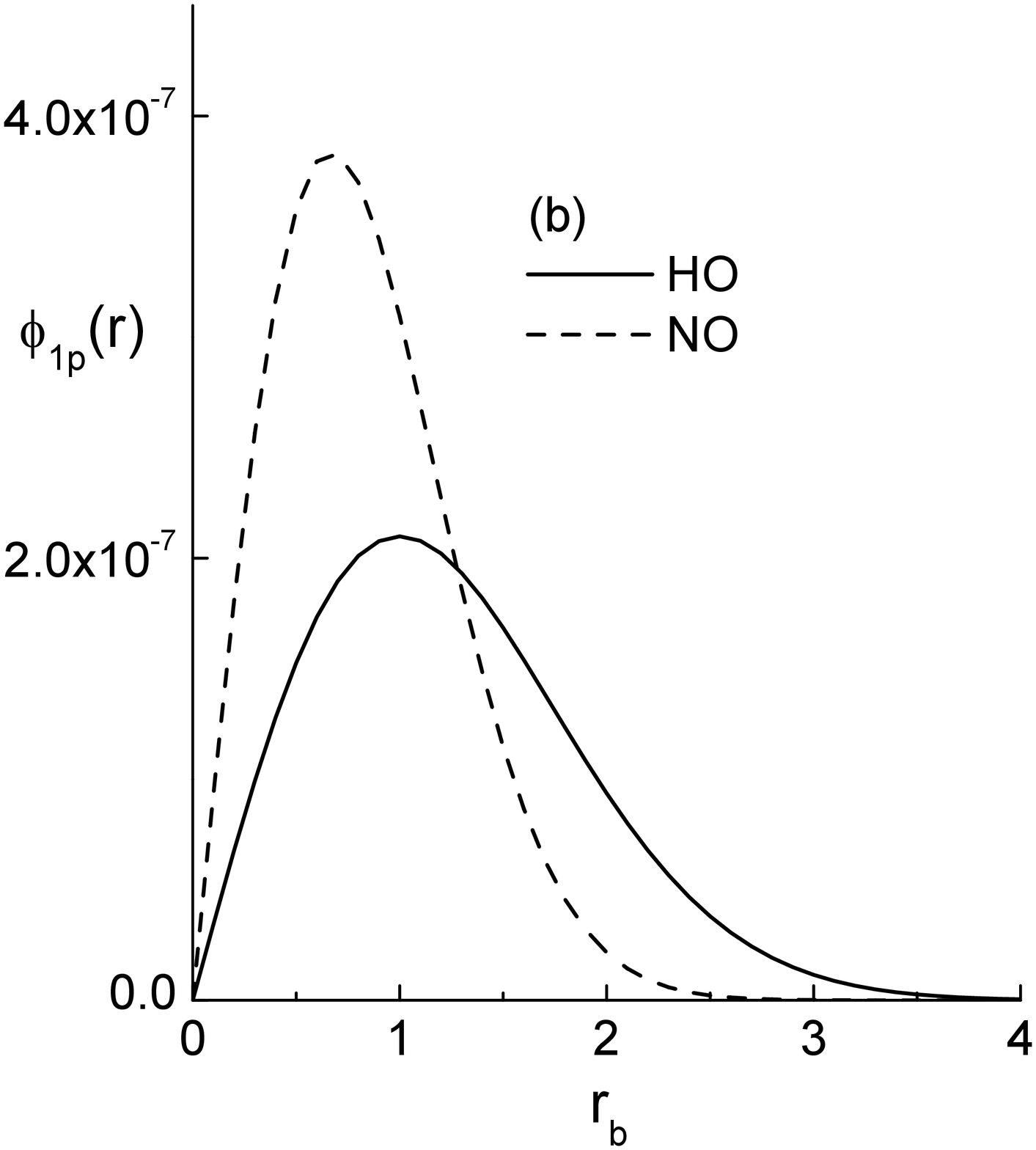}
\hspace{0.3cm}
\includegraphics[height=5.0cm,width=3.5cm]{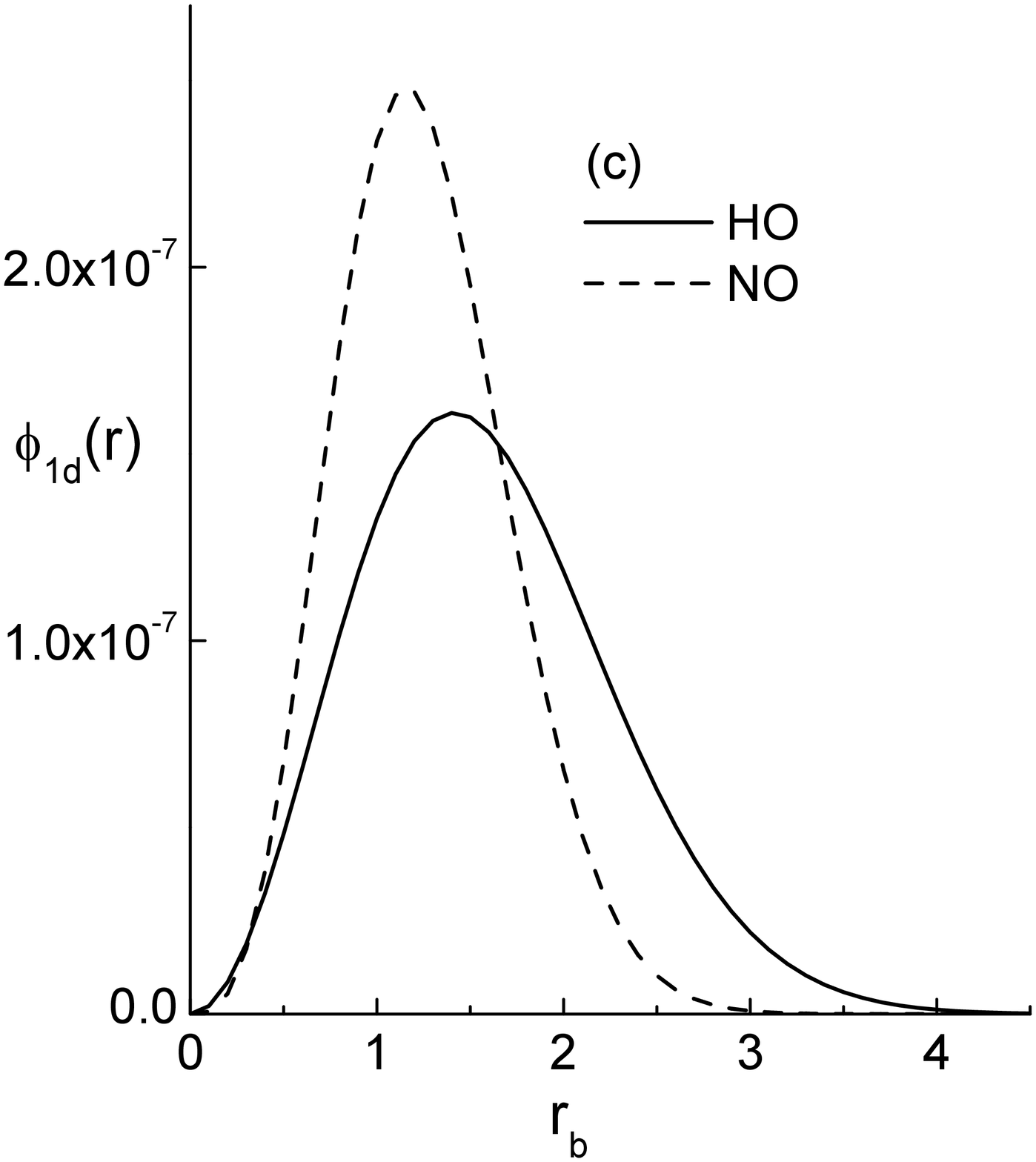}
 \caption{The NO's (dot lines) versus $r_b$ ($r_b=r/b$) of the
states (a) 1s, (b) 1p,  and  (c) 1d obtained by diagonalization of
the one-body density matrix (for $y$=1). The solid lines
correspond to the HO wave-function with the trap length $b=10^4$
\AA.}\label{fig:fig3}
\end{figure}

The NO's of the states $1s$, $1p$ and $1d$ for $y=1$ are shown in
Fig. \ref{fig:fig3}. It is seen that the interatomic correlations
in the $1s$-state spread out the ground state wave function and
consequently the condensation appears in the outer region of the
trap. From the same figure it is obvious that the NO's of the $1p$
and $1d$ states are much more localized in coordinate space than
the equivalent HO orbitals.

\begin{figure}[h]
\centering
\includegraphics[height=5.0cm,width=3.5cm]{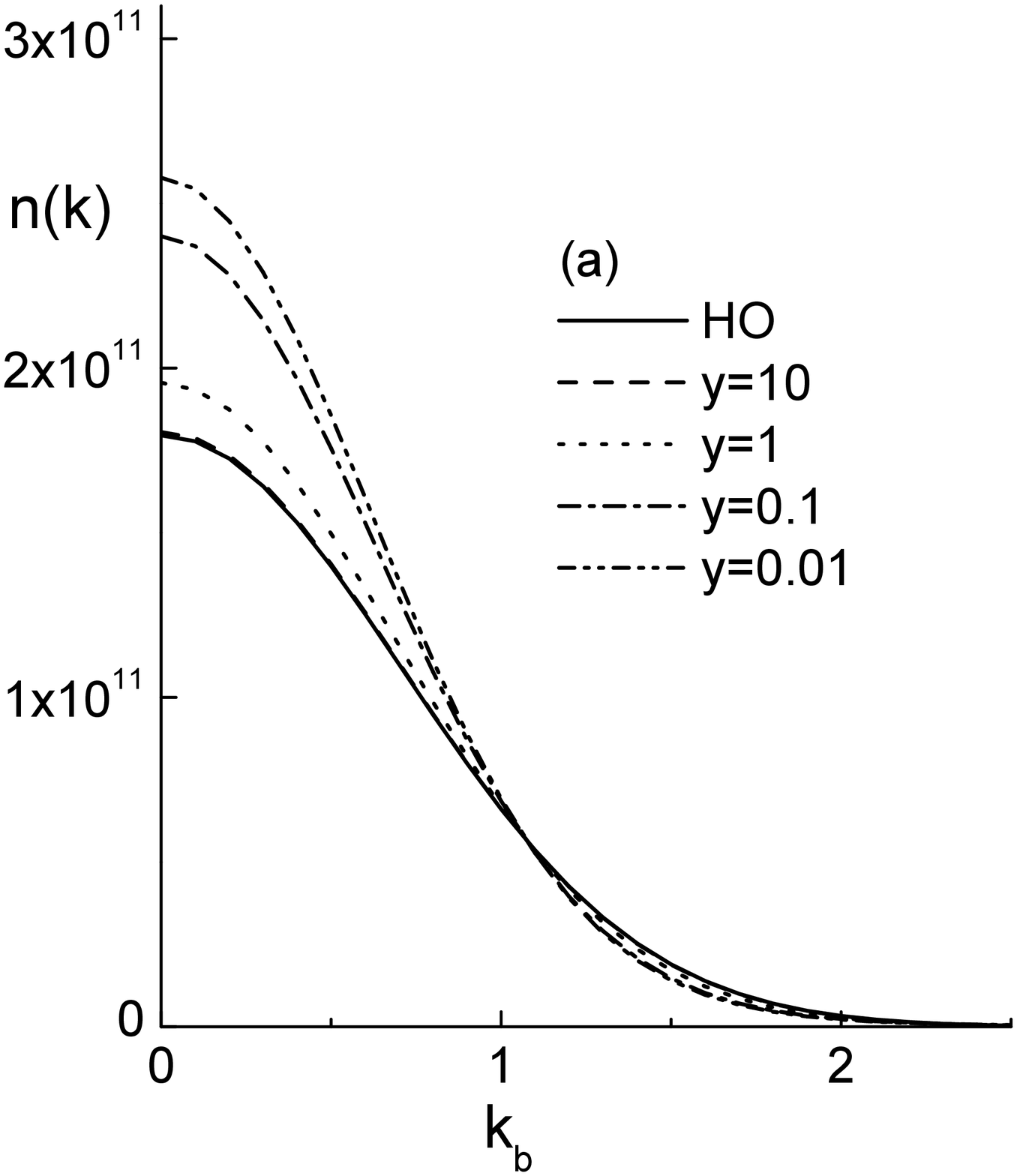}
\hspace{0.3cm}
\includegraphics[height=5.0cm,width=3.5cm]{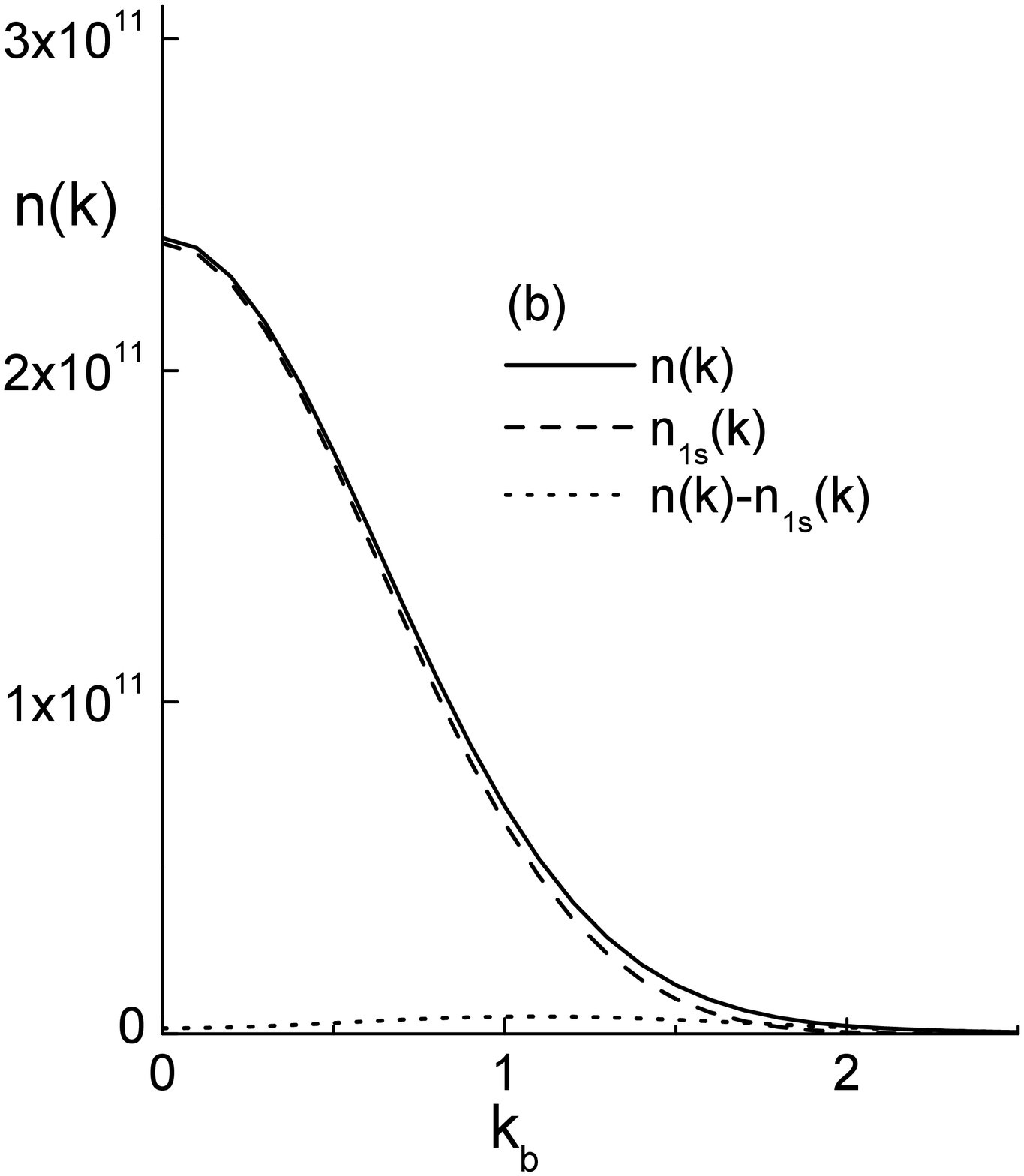}
\hspace{0.3cm}
\includegraphics[height=5.0cm,width=3.5cm]{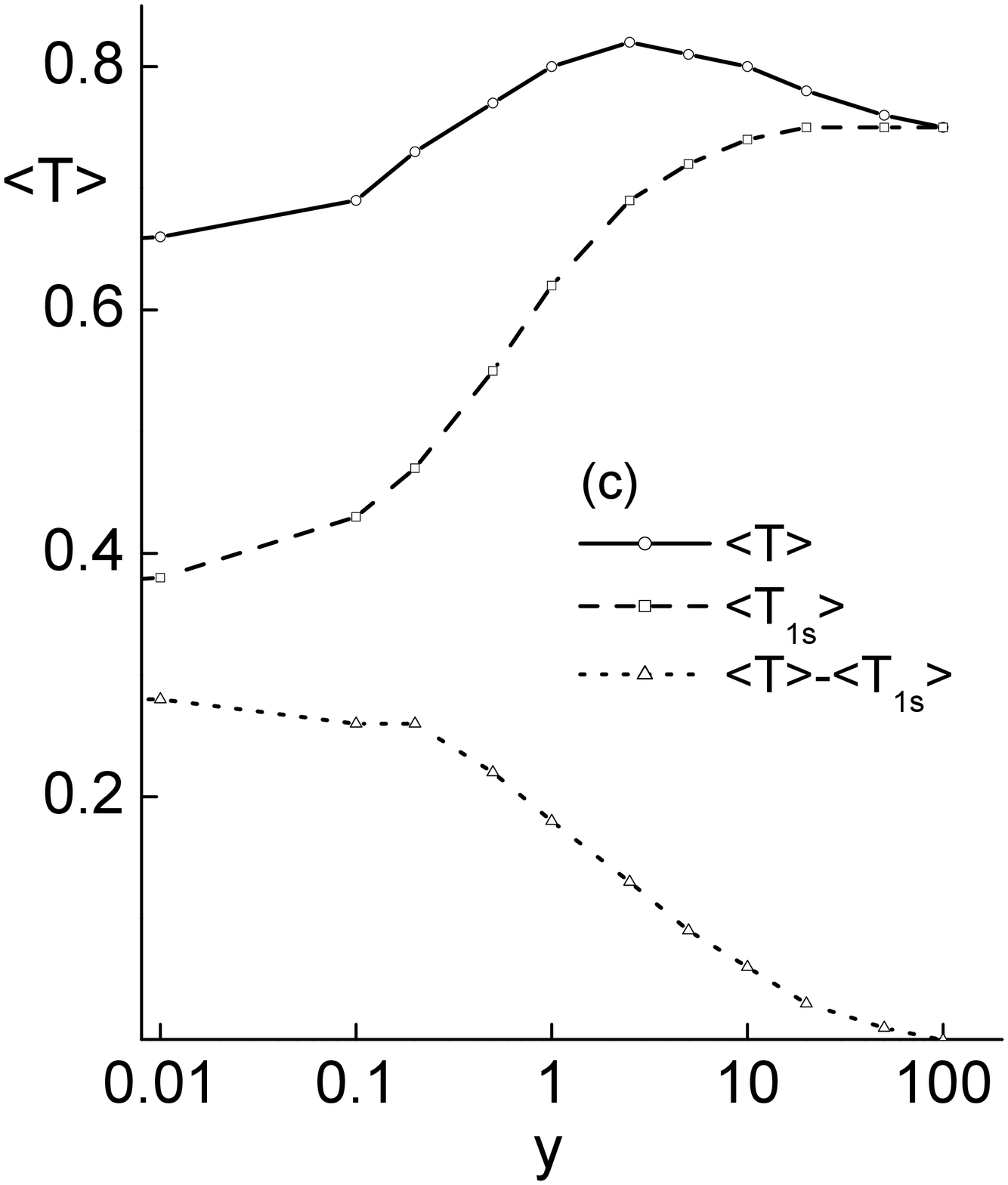}
 \caption{(a) The normalized to $1$ momentum distribution versus
$k_b$ ($k_b=kb$) for various values of the correlation parameter
$y$. (b) The momentum distribution and the contribution to it from
the NO's of the $1s$ NO state and of the rest of the NO states
(for $y=0.1$). (c) The mean kinetic energy per atom, $\langle T_b
\rangle$ ($T_b=T/(\hbar\omega)$), versus $y$. The solid curve
corresponds to the total values of $\langle T_b \rangle$, while
dashed and dotted lines are the contributions to the total
$\langle T_b \rangle$ of the NO's of the $1s$ NO state and of the
rest of the NO states, respectively.}\label{fig:fig4}
\end{figure}

The momentum distribution  can be calculated analytically from Eq.
(\ref{cluster-nk}) or by Fourier transform of the NO's. The
momentum distribution calculated analytically for various values
of the parameter $y$ has been plotted in Fig. \ref{fig:fig4}(a).
It is seen that the large values of $y$ ($y>10$) correspond to the
Gaussian distribution, while when $y$ becomes small enough ($y<1$)
the momentum distribution has a sharp maximum for $k=0$. The
momentum distribution of the $1s$ NO state as well as of the rest
of the NO states for $y=0.1$ are shown and compared with the total
momentum distribution in Fig. \ref{fig:fig4}(b). It is obvious
that although the $1s$ NO state gives the main contribution to the
momentum distribution, the additional NO states contribute to the
momentum distribution mainly in the large values of the momentum
$k$.

The dependence of the mean kinetic energy $\langle T \rangle$ on
the parameter $y$  calculated analytically, using Eq.
(\ref{eq:kinetic-1}), is presented in Fig. \ref{fig:fig4}(c). It
is seen that $\langle T \rangle$ has a maximum for $y\simeq2.5$.
It is interesting to note that for the same value of the parameter
$y$ the NON's of the states $1d$  and $1f$ have the same value as
can be seen from Table \ref{tbl:table1}. The contribution of the
$1s$ NO state and of the rest of the NO states to $\langle T
\rangle$ are shown in the same figure. It is seen that, for large
values of the parameter $y$ (weak correlations) the main
contribution to $\langle T \rangle$ comes from the $1s$ NO state,
while for strong correlations there is a significant contribution
coming from the NO's of the additional states.

A few comments are appropriate. In this section we study the
behavior of various condensate quantities treated in the Jastrow
manner, which introduces one parameter. The determination of that
parameter could be made by fit of the theoretically calculated
quantities (density distribution, momentum distribution,
$\sqrt{\langle r^2\rangle}$, and $\langle T \rangle$) to the
experimental ones as we mentioned in the end of subsection
\ref{sub:sub3-1}, provided that there are experimental data for
the corresponding quantities. It could be determined also by using
the density distribution or the two-body density matrix as a trial
one and applying the variation principle to the ground state
energy of the system. The present approach is quite frequent in
the study of the quantum many body problem when the solution of
the Schr\"odinger equation is very difficult. It should be noted
also that in the present work there is not a direct dependence
between the condensation and the number of the atoms. The
inter-particle correlations are incorporated in the mean field
only by the correlation function which, in some way, depends on
the effective size of the atoms. That dependence can be found from
the information entropy $S$ using the linear dependence of $S$ on
$\ln (Na_b)$ and the linear dependence of $S$ on
$\ln(\frac{1}{y})$ (see Sec. \ref{sub:sub5-1}).

\subsection{Static Structure Factor}

In order to calculate the static structure factor in the framework
of the atomic calculations we choose two trial forms for $g(r)$
\cite{Moustakidis04}. The first one is a gaussian type which has
been extensively and successfully used for the study of similar
problems in atomic physics (Bose gas, liquid helium) as well in
nuclear physics. The relevant $g(r)$ and the entailed $S(k)$ (Case
1) are
\begin{eqnarray}
g(r)&=& 1- \exp[-\beta r^2], \nonumber \\
S(k)&=&1+N(C_1-1)\exp\left[-\frac{k_b^2}{2}\right] \nonumber \\
& &-\frac{N
C_1}{(1+2y^2)^{3/2}}\exp\left[-\frac{k_b^2}{2}(1+2y)\right],
\label{fin-cs-1}
\end{eqnarray}
where $k_b=kb$, $y=\beta b^2 $, $\beta$ is the correlation
parameter and $C_1$ is the normalization factor.

\begin{figure}[h]
\centering
\includegraphics[height=5.0cm,width=3.5cm]{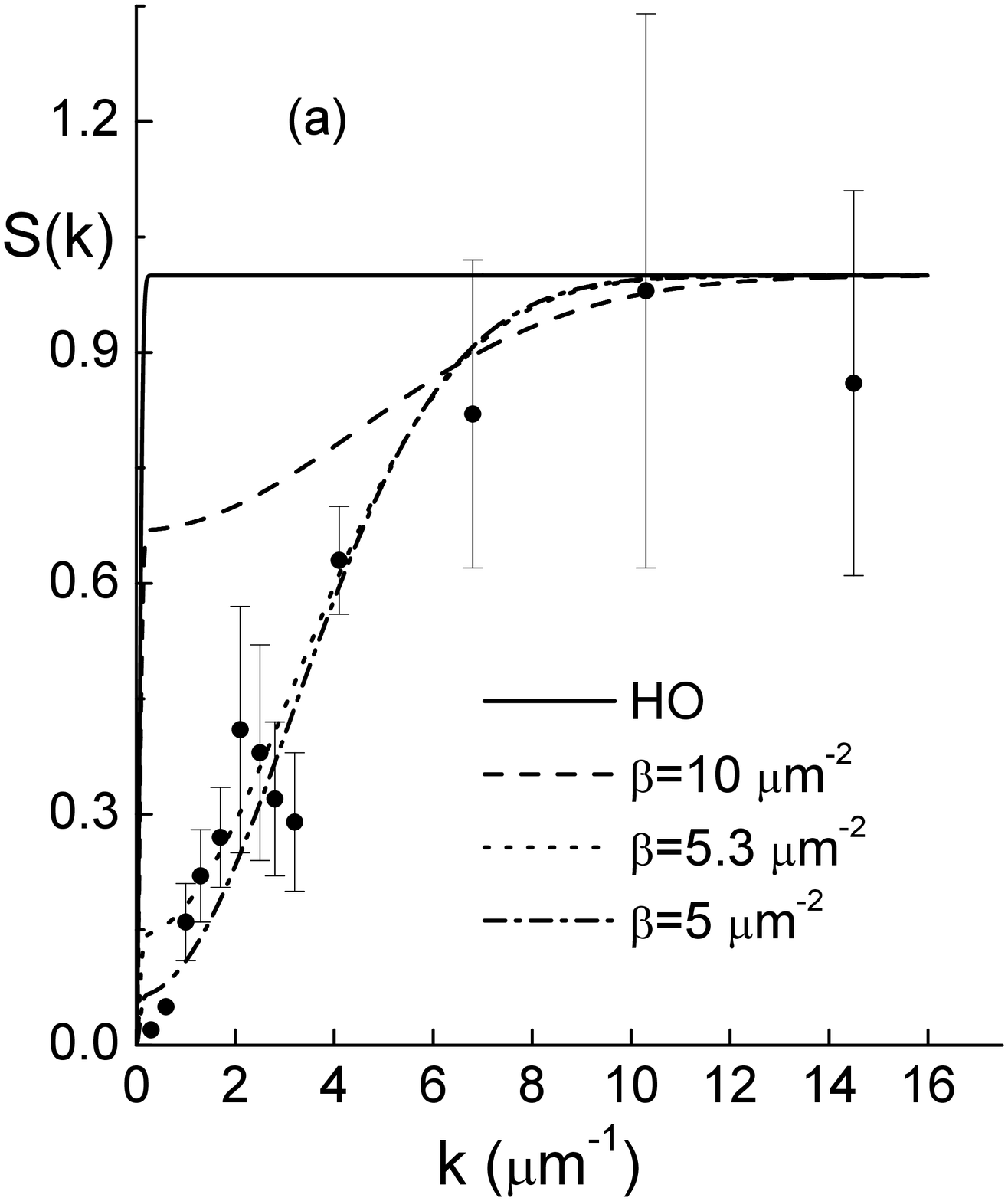}
\hspace{0.3cm}
\includegraphics[height=5.0cm,width=3.5cm]{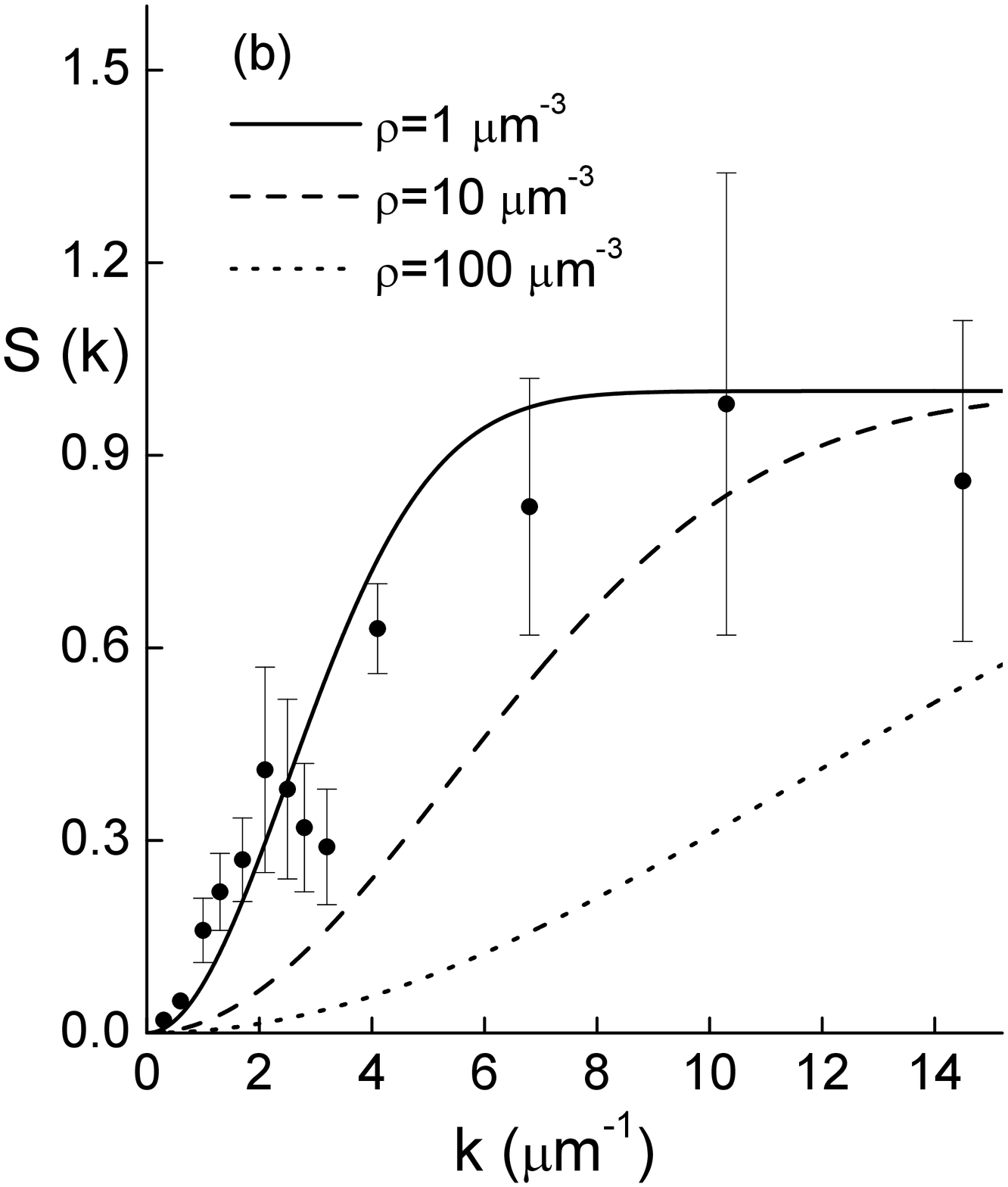}
 \caption{The static structure factor $S(k)$ of the trapped Bose gas
 in various cases versus the momentum $k$,
(a) in Case 1 for various values of the correlation parameter
$\beta$ as well as for the uncorrelated case (harmonic
oscillator), (b) in Case 2 for the least squares best fit value of
the parameter $a$. The experimental points are from reference
\cite{Steinhauer}. For the various cases see
text.}\label{fig:fig5}
\end{figure}

\begin{figure}[h]
\centering
\includegraphics[height=5.0cm,width=3.5cm]{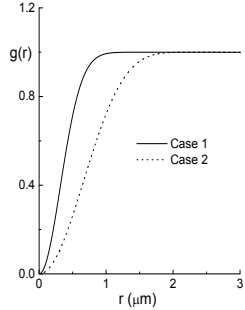}
 \caption{The radial distribution function $g(r)$ for Case 1 and
2 (corresponding to inhomogeneous Bose gas) with the best fit
values of the correlation parameters.}\label{fig:fig6}
\end{figure}

The second trial function $g(r)$ and the relevant $S(k)$ (Case 2)
are of the form
\begin{eqnarray}
g(r)&=& 1- \frac{\sin^4ar}{(ar)^4}, \nonumber \\
S(k)&=&1+N(C_2-1)\exp\left[-\frac{k_b^2}{2}\right] \nonumber\\
& &- \frac{N C_2}{2^{15/2} ab \pi k_b} \sum_{i=1}^{5}
\alpha_i\left[\beta_i\exp\left[-\frac{\beta_i^2}{4}\right]+
\sqrt{\pi}\left(1+\frac{\beta_i^2}{2}\right) {\rm
erf}\left(\frac{\beta_i}{2}\right) \right], \label{fin-cs-2}
\end{eqnarray}
where $a$ is the correlation parameter, $\alpha_i$ are known
coefficients, $\beta_i=\beta_i(a,b,k_b)$, ${\rm
erf}(z)=\displaystyle{\frac{2}{\sqrt{\pi}}} \int_{0}^{z} e^{-t^2}
dt$ and $C_2$ is the normalization factor \cite{Moustakidis04}.

The behavior of S(k), in Case 1, for various values of the
correlation parameter $\beta$ is shown in Fig. \ref{fig:fig5}(a).
It is obvious that the effect of correlations, induced by the
function $g(r)$, becomes large when the parameter $\beta$ becomes
small and vice versa. The case where $\beta\rightarrow \infty$,
corresponds to the uncorrelated case (HO). For the values of $k$
employed in the experiment of Ref. \cite{Steinhauer} (hereafter
EXP) the prediction of the HO model is always close to 1 for
$S(k)$. When the correlation parameter $\beta$  decreases
considerably (strong correlations) the theoretical prediction of
$S(k)$ is in good agreement with the experimental data. The value
$\beta$=5.3 ${\rm \mu m}^{-2}$ gives the best least squares fit in
that case. In general the gaussian form of $g(r)$, in spite of its
simplicity, reproduces fairly well the experimental data of EXP,
both in low and high values of the momentum $k$.  Within our
theoretical model, the gaussian type of $g(r)$ is flexible enough
to obtain values for $S(k)$ in agreement with the experimental
data.

Fig. \ref{fig:fig5}(b) displays the results in Case 2, which are
compared with those of the data of EXP. The model reproduces well
the experimental data in the range 1.5-3 ${\rm \mu m}^{-1}$ (with
best least squares fit value $a=1.34 \  \mu m^{-1}$), but fails in
the range $k>3$ ${\rm \mu m}^{-1}$. The main drawback of this
model is the predicted negative values of  $S(k)$ in the range
close to $k=0$ when the correlation parameter $a$ decreases
considerably (strong correlation case).

The correlation function $g(r)$ corresponding to Cases 1 and 2 for
the correlation parameters $\beta$=5.3 ${\rm \mu m}^{-2}$ and
$a=1.34 \ \mu m^{-1} $ respectively is sketched in Fig.
\ref{fig:fig6}. Those values of the parameters $\beta$ and $a$
give the best $x^2$ value in the fit of the theoretical
expressions of $S(k)$ to the data of EXP. The most striking
feature in Case 2 is the existence of strong correlations,
introduced by  $g(r)$, in order to reproduce the experimental data
of $S(k)$. It is worthwhile to point out that $g(r)$, in Case 2,
exhibits fluctuations in the range $r>2$ ${\rm \mu m}$ but this is
not visible in Fig. \ref{fig:fig6}.

The possibility of a linear dependence of $S(k)$ on $k$ for small
values of $k$, as predicted from other works \cite{Zambelli}, is
prohibitive, on the basis of Eq. (\ref{eq:eq18}) at least in the
case where the trap is an harmonic oscillator one. That can be
seen considering  the ground state wave function to be the
harmonic oscillator one and transforming ${\bf r}_1$ and ${\bf
r}_2$ in Eq. (\ref{eq:eq18}) into the coordinates of the relative
motion (${\bf r}={\bf r}_1-{\bf r}_2$) and the center of mass
motion (${\bf R}=({\bf r}_1+{\bf r}_2)/2 $). After some algebra
$S({\bf k})$ takes the form
\begin{equation}
S({\bf k}) \sim \int \textrm{exp} \left[ i{\bf k}{\bf r} \right]
 \textrm{exp}\left[ -r_b^2\right] [Cg(r)-1] d {\bf r}. \label{sk-rel}
\end{equation}

For finite systems, as a trapped Bose gas, we can expand the
exponential $\textrm{exp} \left[ i{\bf k}{\bf r} \right]$, since
${\bf r}$ is bounded. Thus:
\begin{equation}
\textrm{exp} \left[ i{\bf k}{\bf r} \right]=1+i{\bf k}{\bf r}
+\frac{(i{\bf k}{\bf r})^2}{2 !}+ \frac{(i{\bf k}{\bf r})^3}{3
!}+\cdots .\label{expand}
\end{equation}

Substituting  Eq. (\ref{expand}) into Eq. (\ref{sk-rel}) and
taking into account that the terms with odd powers of $k$ do not
contribute to the integral, $S(k)$ takes the form
\begin{equation}
S(k) \sim  a_1k^2+a_2k^4+\cdots .\label{sk-rel-2}
\end{equation}

Hence, for small values of $k$, $S(k)$ depends linearly on $k^2$.
The gaussian factor $\textrm{exp}\left[ -r_b^2\right]$,
originating from the harmonic oscillator wave function of the
trapped Bose gas, ensures the convergence of the integrals $a_i$
corresponding to the even powers of the expansion
\cite{Moustakidis04}.

\section{Quantum-Information properties of trapped
Bose gas}\label{sec:sec4}

\subsection{Shannon Information Entropy}

The Boltzmann-Gibbs-Shannon information entropy
\cite{Shannon48,Halliwell93} of a finite probability distribution
($p_1$,$p_2$,$\cdot$,$p_k$) is defined as the quantity
\begin{equation}
S=-\sum_{i=1}^{k}p_i \ln p_i, \label{BGS}
\end{equation}
with the constraint: $\displaystyle{\sum_{i=1}^{k}p_i=1}$. $S$ is
measured in bits if the base of the logarithm is 2 and nats
(natural units of information) if the logarithm is natural.

$S$ appears in different areas: information theory, ergodic theory
and statistical mechanics. It is closely related to the entropy
and disorder in thermodynamics. The maximum value of $S$ is
obtained if $p_1=p_2=\cdots=p_k=\frac{1}{k}$ i.e. $S_{max}=\ln k$.
The minimum value of $S$ is found when one of the $p_i$'s is equal
$1$ and all the others are equal to $0$. Then $S_{min}=0$. The
above definition holds for discrete probability distributions
\cite{Chatzisavvas05}. In quantum mechanics we are often
interested in a continuous probability distribution $p(x)$. In
this case the obvious generalization of Eq. (\ref{BGS}) is the
information entropy
\begin{equation}
S=-\int p(x) \ln p(x) dx, \label{S-con}
\end{equation}
where $\int p(x) dx=1$. Now $p(x)$ is a quantum mechanical
probability distribution and $S$ may be called the quantum entropy
\cite{Ohya93}. $S$ indicates the amount of disorder or randomness
(uncertainty) in a physical system. Shannon considered this
uncertainty attached to the system as the amount of information
carried by the system. If a physical system has  a large
uncertainty and one obtains information on the system by some
procedure, as a measurement, then the information is more valuable
than that received from a system having less uncertainty. Thus,
before a measurement, the uncertainty of the position of a
particle is small for a localized probability distribution, while
for a diffuse distribution is large. The same holds for the
missing information due to a limited knowledge of the system via a
probability distribution. After the measurement the gain in
information for a localized distribution is smaller than the
corresponding gain for a diffuse distribution.

An important step is the discovery of an entropic uncertainty
relation (EUR) \cite{Bialynicki75}, which for a three-dimensional
system has the form
\begin{equation}\label{eq:equ2}
    S=S_r+S_k\geq 3\,(1+\ln{\pi})\simeq 6.434,
\end{equation}
where $S_r$ is the information entropy in position-space of the
density distribution $\rho(\textbf{r})$ of a quantum system
\begin{equation}\label{eq:equ3}
    S_r=-\int
    \rho(\textbf{r})\,\ln{\rho(\textbf{r})}\,d\textbf{r},
\end{equation}
and $S_k$ is the information entropy in momentum-space of the
corresponding momentum distribution $n(\textbf{k})$
\begin{equation}\label{eq:equ4}
    S_k=-\int n(\textbf{k})\,\ln{n(\textbf{k})}\,d\textbf{k}.
\end{equation}

The total information entropy is given by
\begin{equation}\label{eq:stot}
    S=S_r+S_k.
\end{equation}

The density distributions $\rho(\textbf{r})$ and $n(\textbf{k})$
are normalized to one. Inequality (\ref{eq:equ2}), for the
information entropy sum in conjugate spaces, is a joint measure of
uncertainty of a quantum mechanical distribution, since a highly
localized $\rho(\textbf{r})$ is associated with a diffuse
$n(\textbf{k})$, leading to low $S_r$ and high $S_k$ and
vice-versa. Expression (\ref{eq:equ2}) is an
information-theoretical relation stronger than Heisenberg's.

In previous work we proposed a universal property of $S$ for the
density distributions of nuclei, electrons in atoms and valence
electrons in atomic clusters \cite{Massen98}. This property has
the form
\begin{equation}\label{eq:equ5}
    S=a+b \ln{N},
\end{equation}
where $N$ is the number of particles of the system and the
parameters $a, b$ depend on the system under consideration. It is
noted that recently we have obtained the same form for systems of
correlated bosons in a trap \cite{Massen02}. This concept was also
found to be useful in a different context. Using the formalism in
phase-space of Ghosh, Berkowitz and Parr \cite{Ghosh84}, we found
that the larger the information entropy, the better the quality of
the nuclear density distribution \cite{Lalazissis98}. Recently the
Shannon information entropy has been applied successfully to the
study of the free expansion of impenetrable bosons on the
one-dimensional optical lattices \cite{Rigol05}.

\subsection{Onicescu's Information Entropy}

Onicescu tried to define a finer measure of dispersion
distributions than that of Shannon's information entropy
\cite{Onicescu96}. Thus, he introduced the concept of information
energy $E$. For a discrete probability distribution
$(p_1,p_2,\ldots,p_k)$ the information energy $E$ is defined by
\begin{equation}\label{eq:equ9}
    E=\sum_i^k p_i^2,
\end{equation}
which is extended for a continuous density distribution $\rho(x)$
as
\begin{equation}\label{eq:equ10}
    E=\int \rho^2(x)\,dx.
\end{equation}
The meaning of (\ref{eq:equ10}) can be seen by the following
simple argument: For a Gaussian distribution of mean value $\mu$,
standard deviation $\sigma$ and normalized density
\begin{equation}\label{eq:equ11}
  \rho(x)=\frac{1}{\sqrt{2\pi}\sigma}\, \textrm{exp}
  \left[-\frac{(x-\mu)^2}{2\sigma^2} \right],
\end{equation}
relation (\ref{eq:equ10}) gives
\begin{equation}\label{eq:equ12}
  E=\frac{1}{2\pi\sigma^2} \int_{-\infty}^{\infty}
   \textrm{exp}
  \left[-\frac{(x-\mu)^2}{\sigma^2}
  \right]\,dx=\frac{1}{2\sigma\sqrt{\pi}}.
\end{equation}
$E$ is maximum if one of the $p_i$'s equals 1 and all the others
are equal to zero i.e. $E_{max}=1$, while $E$ is minimum when
$p_1=p_2=\ldots=p_k=\frac{1}{k}$, hence $E_{min}=\frac{1}{k}$
(total disorder). $E$ has been called information energy, although
it does not have the dimension of energy \cite{Lepadatu03}. This
is due to the fact that $E$ becomes minimum for equal
probabilities (total disorder), by analogy with thermodynamics.

It is seen from (\ref{eq:equ12}) that the greater the information
energy, the more concentrated is the probability distribution,
while the information content decreases. $E$ and information
content are reciprocal, hence one can define the quantity
\cite{MoustaChatz05}
\begin{equation}\label{eq:equ13}
  O=\frac{1}{E},
\end{equation}
as a measure of the information content of a quantum system
corresponding to Onicescu's information energy.

Relation (\ref{eq:equ10}) is extended for a 3-dimensional
spherically symmetric density  and momentum distribution as follow
$\rho(\textbf{r})$
\begin{eqnarray}\label{eq:equ14}
  E_r=\int \rho^2(\textbf{r})\,d\textbf{r} \nonumber \\
  E_k=\int n^2(\textbf{k})\,d\textbf{k}.
\end{eqnarray}

$E_r$ has dimension of inverse volume, while $E_k$ of volume. Thus
the product $E_r E_k$ is dimensionless and can serve as a measure
of concentration (or information content) of a quantum system. It
is also seen from (\ref{eq:equ12}),(\ref{eq:equ13}) that $E$
increases as $\sigma$ decreases (or concentration increases) and
the information (or uncertainty) decreases. Thus $O$ and $E$ are
reciprocal. In order to be able to compare $O$ with Shannon's
entropy $S$, we redifine $O$ as
\begin{equation}\label{eq:equ15}
  O=\frac{1}{E_r E_k},
\end{equation}
as a measure of the information content of a quantum system in
both position and momentum spaces, inspired by Onicescu's
definition.

\subsection{Landsberg's Order Parameter}

Landsberg \cite{Landsberg84} defined the order parameter $\Omega$
(or disorder $\Delta$) as
\begin{equation}
 \Omega = 1-\Delta = 1- \frac{S}{S({\rm max})},
 \label{omega}
 \end{equation}
where $S$ is the information entropy (actual) of the system and
$S({\rm max})$ the maximum entropy accessible to the system. Thus
the concepts of entropy and disorder are decoupled and it is
possible for the entropy and order to increase simultaneously. It
is noted that $\Omega =1$ corresponds to perfect order and
predictability, while $\Omega =0$ means complete disorder and
randomness.

\subsection{Two-body information entropies}

The two-body Shannon information entropy both in position- and
momentum-space and in total are defined respectively
\cite{MoustaChatz05,Amovilli04,Cover91}
\begin{equation}
S_{2r}=-\int \rho({\bf r}_1,{\bf r}_2) \ln \rho({\bf r}_1,{\bf
r}_2) d {\bf r}_1 d{\bf r}_2 \label{S2r-1}
\end{equation}
\begin{equation}
S_{2k}=-\int n({\bf k}_1,{\bf k}_2) \ln n({\bf k}_1,{\bf k}_2)
d{\bf k}_1 d{\bf k}_2, \label{S2k-1}
\end{equation}
\begin{equation}
S_2=S_{2r}+S_{2k}. \label{S2-1}
\end{equation}

The one-body Onicescu's information entropy is already defined in
(\ref{eq:equ14}) and (\ref{eq:equ15}), where the generalization to
the two-body information entropy is straightforward and is given
by
\begin{equation}
O_2= \frac{1} {E_{2r} E_{2k}}, \label{O1-2}
\end{equation}
where
\begin{eqnarray}
E_{2r}&=&\int \rho^2({\bf r}_1,{\bf r}_2) d{\bf r}_1 d{\bf r}_2 \nonumber \\
E_{2k}&=&\int n^2({\bf k}_1,{\bf k}_2) d{\bf k}_1 d {\bf k}_2.
\label{E2r-2k}
\end{eqnarray}

\subsection{Kullback-Leibler relative entropy and Jensen-Shannon
divergence}

A well known measure of distance of two discrete probability
distributions $p_i^{(1)}, p_i^{(2)}$ is the Kullback-Leibler
relative entropy \cite{Kullback59}
\begin{equation}
    K(p_i^{(1)},p_i^{(2)})=\sum_i
    p_i^{(1)}\,\ln{\frac{p_i^{(1)}}{p_i^{(2)}}}, \label{eq:equ6}
\end{equation}
which for continuous probability distributions $\rho^{(1)},
\rho^{(2)}$ is defined as
\begin{equation}
    K=\int
    \rho^{(1)}(x)\,\ln{\frac{\rho^{(1)}(x)}{\rho^{(2)}(x)}}\,dx, \label{eq:equ7}
\end{equation}
which can be easily extended for 3-dimensional systems.

Our aim is to calculate the relative entropy (distance) between
$p^{(1)}$ (correlated) and $p^{(2)}$ (uncorrelated) densities both
at the one- and the two-body levels in order to assess the
influence of short range correlations (SRC) through the
correlation parameter $y$, on the distance $K$
\cite{MoustaChatz05}. It is noted that this is done for both
systems under consideration: nuclei and trapped Bose gases. An
alternative definition of distance of two probability
distributions was introduced by Rao and Lin \cite{Rao87,Lin91},
i.e. a symmetrized version of $K$, the Jensen-Shannon divergence
$J$ \cite{Majtey05}
\begin{equation}
    J(p^{(1)},p^{(2)})=H\left(\frac{p^{(1)}+p^{(2)}}{2}\right)-\frac{1}{2}H\left(p^{(1)}\right)
    -\frac{1}{2}H\left(p^{(2)}\right), \label{eq:equ8}
\end{equation}
where $H(p)=\displaystyle{-\sum_i p_i \ln{p_i}}$ stands for
Shannon's entropy. We expect for strong SRC the amount of
distinguishability of the correlated from the uncorrelated
distributions is larger than the corresponding one with small SRC.
We may also see the effect of SRC on the number of trials $L$
needed to distinguish $p^{(1)}$ and $p^{(2)}$ (in the sense
described in \cite{Majtey05}).

In addition to the above considerations, we connect $S_r$ and
$S_k$ with fundamental quantities i.e. the root mean square radius
and kinetic energy respectively. We also argue on the effect of
SRC on EUR and we propose a universal relation for $S$, by
extending our formalism from the one- and two-body level to the
$N$-body level, which holds exactly for uncorrelated densities in
trapped Bose gas and it is conjectured to hold approximately for
correlated densities in  Bose gases (see Sec. \ref{sub:sec4-7}).

The Kullback-Leibler relative information entropy $K$ for
continuous distributions $\rho_i^{(1)}$ and $\rho_i^{(2)}$ is
defined by relation (\ref{eq:equ7}). It measures the difference of
$\rho_i^{(1)}$ form the reference (or apriori) distribution
$\rho_i^{(2)}$. It satisfies: $K\geq 0$ for any distributions
$\rho_i^{(1)}$ and $\rho_i^{(2)}$. It is a measure which
quantifies the distinguishability (or distance) of $\rho_i^{(1)}$
from $\rho_i^{(2)}$, employing a well-known concept in standard
information theory. In other words it describes how close
$\rho_i^{(1)}$ is to $\rho_i^{(2)}$ by carrying out observations
or coin tossing, namely $L$ trials (in the sense described in
\cite{Majtey05}). We expect for strong SRC the amount of
distinguishability of the correlated $\rho_i^{(1)}$ and the
uncorrelated distributions $\rho_i^{(2)}$ is larger than the
corresponding one with small SRC.

However, the distance $K$ does not satisfy the triangle inequality
and in addition is i) not symmetric ii) unbounded and iii) not
always well defined \cite{Majtey05}. To avoid these difficulties
Rao and Lin \cite{Rao87,Lin91} introduced a symmetrized version of
$K$ (recently discussed in \cite{Majtey05}), the Jensen-Shannon
divergence $J$ defined by relation (\ref{eq:equ8}). $J$ is minimum
for $\rho^{(1)}=\rho^{(2)}$ and maximum when $\rho^{(1)}$ and
$\rho^{(2)}$ are two distinct distributions, when $J=\ln{2}$. In
our case  $J$ can be easily generalized for continuous density
distributions. For $J$ minimum the two states represented by
$\rho^{(1)}$ and $\rho^{(2)}$ are completely indistinguishable,
while for $J$ maximum they are completely distinguishable. It is
expected that for strong SRC the amount of distinguishability can
be further examined by using Wooter's criterion \cite{Majtey05}.
Two probability distributions $\rho^{(1)}$ and $\rho^{(2)}$ are
distinguishable after $L$ trials $(L\rightarrow \infty)$ if and
only if $\left( J(\rho^{(1)},\rho^{(2)})
\right)^{\frac{1}{2}}>\frac{1}{\sqrt{2L}}$.

The relative entropy is a measure of distinguishability or
distance of two states. It is defined \cite{MoustaChatz05},
generalizing (\ref{eq:equ7}), by
\begin{equation}
    K=\int \psi^2(\textbf{r})
    \ln{\frac{\psi^2(\textbf{r})}{\phi^2(\textbf{r})}}\,d\textbf{r}. \label{eq:equ16}
\end{equation}
In our case $\psi(\textbf{r})$ is the correlated case and
$\phi(\textbf{r})$ the uncorrelated one. Thus
\begin{equation}
    K_{1r}=\int
    \rho(\textbf{r})\,\ln{\frac{\rho(\textbf{r})}{\rho'(\textbf{r})}} \label{eq:equ17}
    \,d\textbf{r},
\end{equation}
where $\rho(\textbf{r})$ is the correlated one-body density and
$\rho'(\textbf{r})$ is the uncorrelated one-body density.

A corresponding formula holds in momentum-space
\begin{equation}
     K_{1k}=\int
     n(\textbf{k})\,\ln{\frac{n(\textbf{k})}{n'(\textbf{k})}}
     \,d\textbf{k},
     \label{eq:equ18}
\end{equation}
where $n(\textbf{k})$ is the correlated one-body density and
$n'(\textbf{k})$ is the uncorrelated one.

For the two-body case we have
\begin{equation}
    K_{2r}=\int \rho(\textbf{r}_1,\textbf{r}_2)\,
    \ln{\frac{\rho(\textbf{r}_1,\textbf{r}_2)}{\rho'(\textbf{r}_1,\textbf{r}_2)}}\,
    d\textbf{r}_1 d\textbf{r}_2, \label{eq:equ19}
\end{equation}
where $\rho(\textbf{r}_1,\textbf{r}_2)$ is the correlated two-body
density in position-space and $\rho'(\textbf{r}_1,\textbf{r}_2)$
is the uncorrelated one.

The generalization to momentum-space is straightforward
\begin{equation}
    K_{2k}=\int n(\textbf{k}_1,\textbf{k}_2)
    \ln{\frac{n(\textbf{k}_1,\textbf{k}_2)}{n'(\textbf{k}_1,\textbf{k}_2)}}\,
    d\textbf{k}_1 d\textbf{k}_2, \label{eq:equ20}
\end{equation}
where $n(\textbf{k}_1,\textbf{k}_2)$ is the correlated two-body
density in momentum-space and $n'(\textbf{k}_1,\textbf{k}_2)$ is
the uncorrelated one.

For the Jensen-Shannon divergence $J$ we may write formulas for
$J_1$ (one-body) and $J_2$ (two-body), employing definition
(\ref{eq:equ8}) and putting the corresponding correlated
$\rho^{(1)}$ and uncorrelated $\rho^{(2)}$ distributions in
position- and momentum-spaces. We calculate $K$ and $J$ in
position- and momentum-spaces, for nuclei and bosons.

\subsection{Numerical Results and Discussion}

For the sake of symmetry and simplicity we put the width of the HO
potential $b=1$. Actually for $b=1$ in the case of uncorrelated
case it is easy to see that $S_{1r}=S_{1k}$ and also
$S_{2r}=S_{2k}$ (the same holds for Onicescu entropy), while when
$b\neq 1$ there is a shift of the values of $S_{1r}$ and $S_{1k}$
by an additive factor $\ln{b^3}$. However, the value of $b$ does
not affect directly the total information entropy $S$ (and also
$O$). $S$ and $O$ are just functions of the correlation parameter
$y$ \cite{MoustaChatz05}.

\begin{figure}[h]
 \centering
 \includegraphics[height=5.0cm,width=3.5cm]{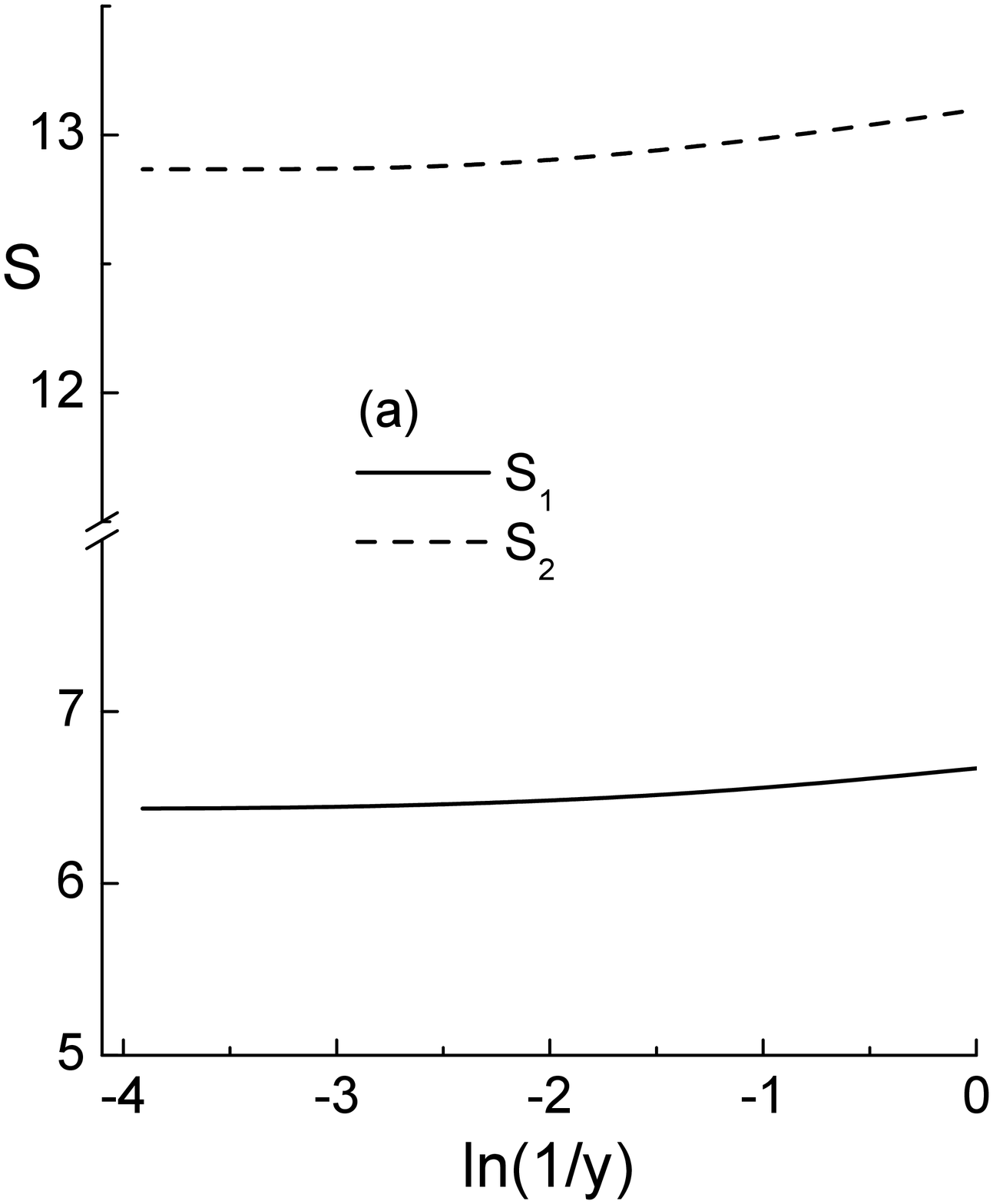}
 \hspace{0.3cm}
 \includegraphics[height=5.0cm,width=3.5cm]{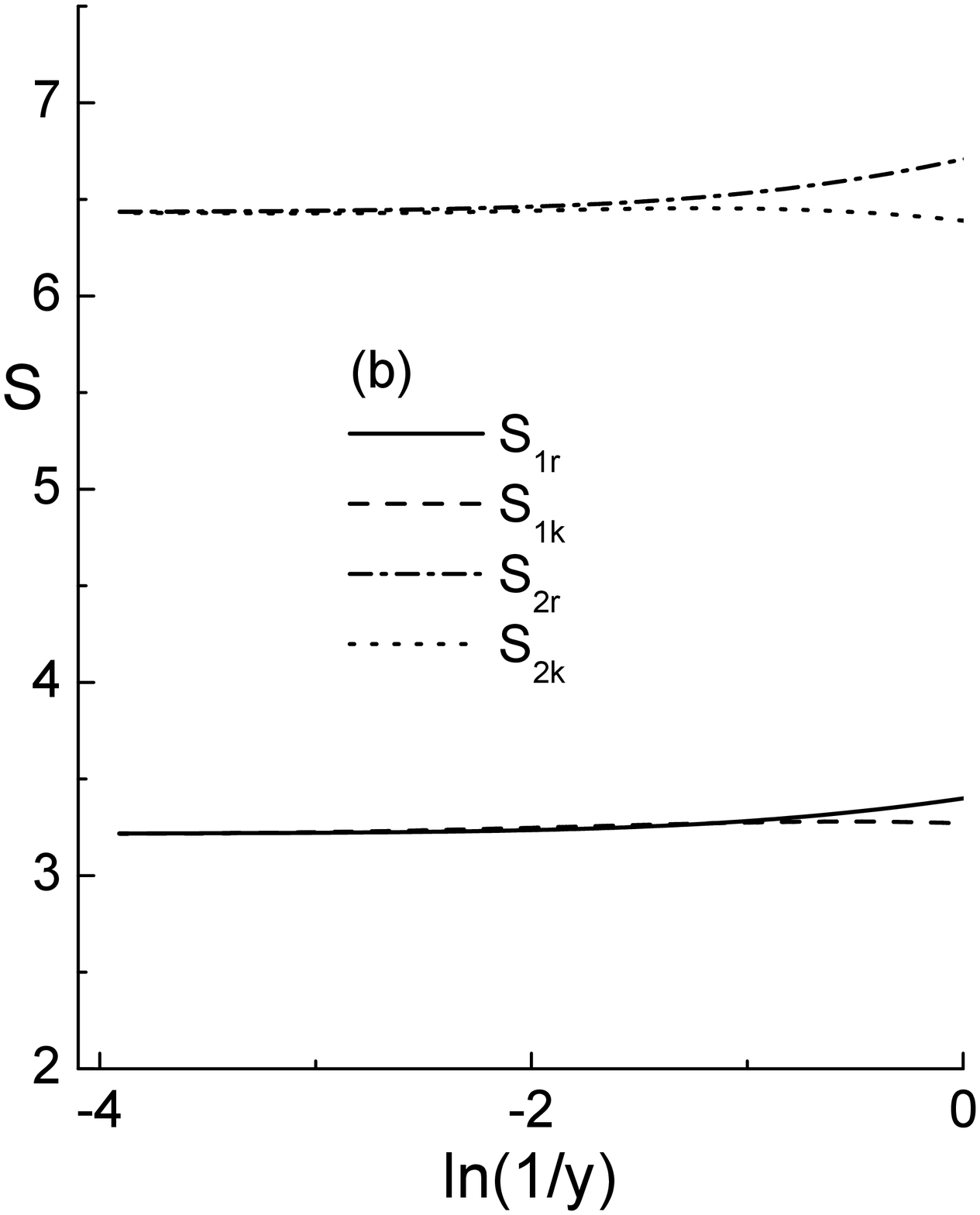}
 \caption{ (a) Shannon information entropy (one- and
 two-body) (b) Shannon information entropy (one- and two-body) both in
coordinate- and momentum-space, in a trapped Bose
gas.}\label{fig:fig7}
\end{figure}

In Fig. \ref{fig:fig7} we present the Shannon information entropy
$S_1$ using relation (\ref{eq:stot}) and $S_2$ using relation
(\ref{S2-1}) in  trapped Bose gas as functions of the correlation
parameter $\ln{(\frac{1}{y})}$. It is seen that $S_1$ and $S_2$
increase almost linearly with the strength of SRC i.e.
$\ln{(\frac{1}{y})}$ in both systems. The relations $S_2=2 S_1$
and $O_2=O_1^2$ hold exactly for the uncorrelated densities while
the above relations are almost exact for the uncorrelated
densities. For the sake of comparison we also present the
decomposition of $S$ in coordinate and momentum spaces i.e.
$S_{1r}$, $S_{1k}$, $S_{2r}$, $S_{2k}$ employing (\ref{eq:equ3}),
(\ref{eq:equ4}), (\ref{S2r-1}), (\ref{S2k-1}). The most striking
feature concluded from the above Figures is the similar behavior
between $S_{1r}$ and $S_{2r}$ and also $S_{1k}$ and $S_{2k}$
respectively.

\begin{figure}
 \centering
 \includegraphics[height=5.0cm,width=3.5cm]{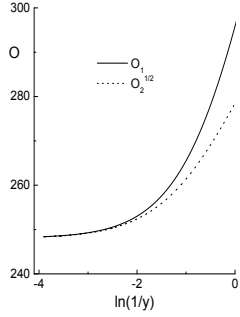}
 \caption{The Onicescu information entropy (both one- and two-body) in a trapped Bose gas.}\label{fig:fig8}
\end{figure}

In Fig. \ref{fig:fig8} we plot the Onicescu information entropy
both one-body $(O_1)$ and two-body $(O_2)$ (relations
(\ref{eq:equ15}), (\ref{O1-2})). We conclude by noting once again
the strong similarities of the behavior between one- and two-body
Onicescu entropy.

\begin{figure}[htb]
 \centering
 \includegraphics[height=5.0cm,width=3.5cm]{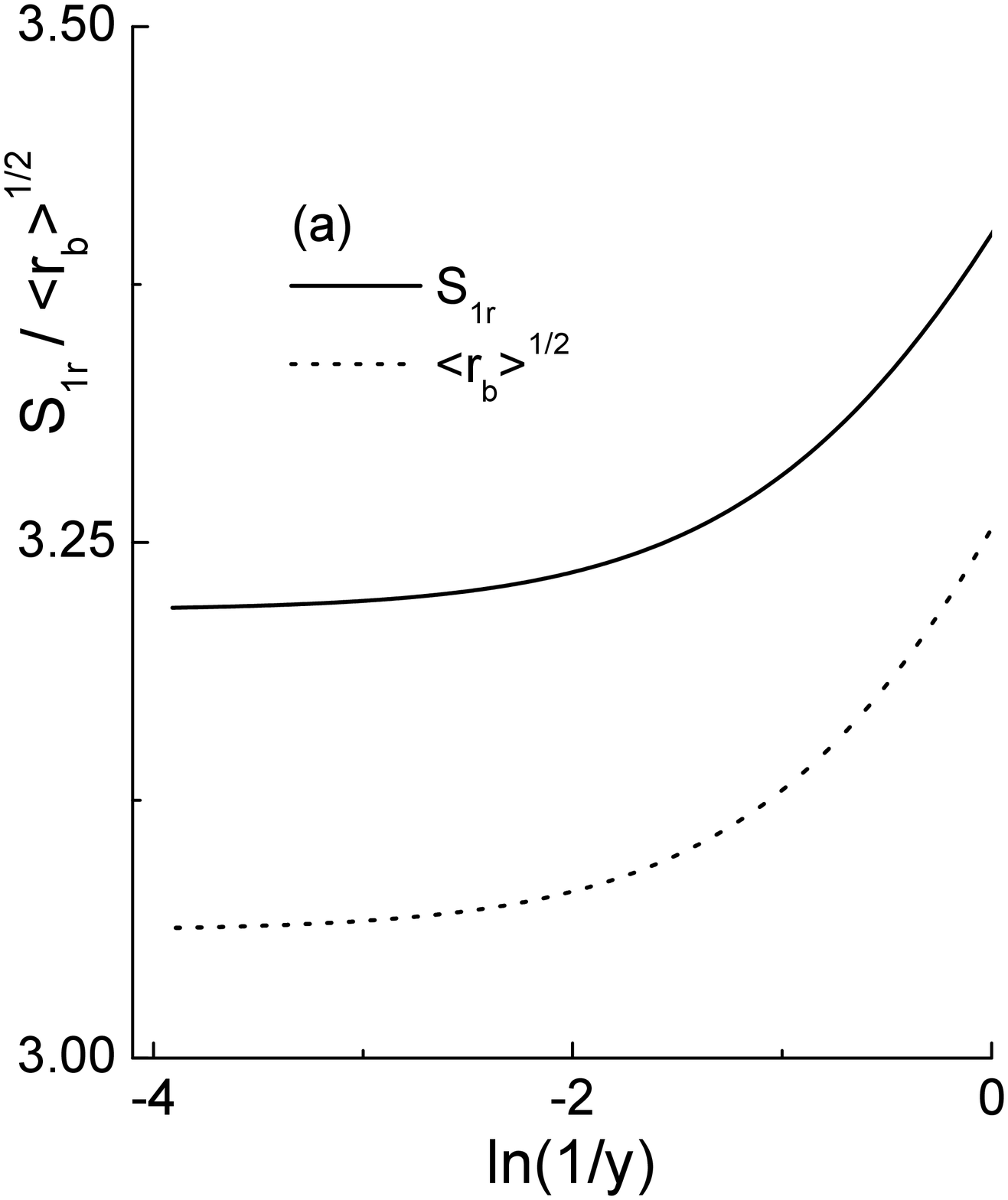}
 \hspace{0.3cm}
 \includegraphics[height=5.0cm,width=3.5cm]{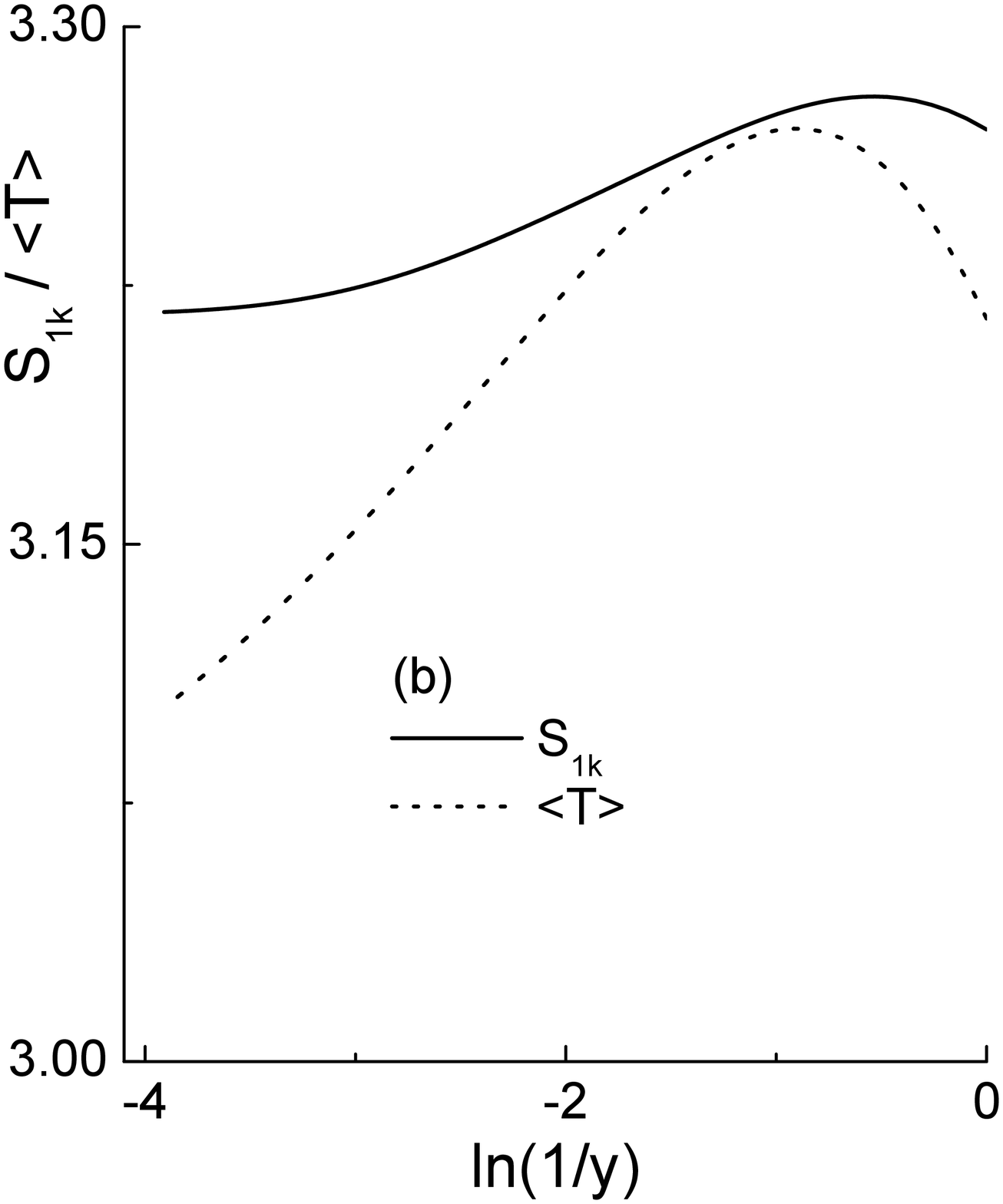}
 \caption{ (a) Mean-square radius and the Shannon information entropy
$S_{1r}$ (b) Mean kinetic energy $\langle T \rangle$ (in $\hbar
\omega$ units) and the Shannon information entropy $S_{1k}$, as
functions of the correlation parameter $\ln{(\frac{1}{y})}$ in a
trapped Bose gas.}\label{fig:fig9}
\end{figure}

It is interesting to observe the relation of the rms radii
$\sqrt{\langle r^2 \rangle}$ with $S_r$ as well as the
corresponding relation of the mean kinetic energy $\langle T
\rangle$ with $S_k$, as functions of the strength of SRC,
$\ln{(\frac{1}{y})}$. This is done in Fig. \ref{fig:fig9} for
$\sqrt{\langle r^2 \rangle}$ and $\langle T \rangle$ after
applying the suitable rescaling. The corresponding curves are
similar for nuclei and trapped Bose gas.

\begin{figure}[htb]
 \centering
 \includegraphics[height=5.0cm,width=3.5cm]{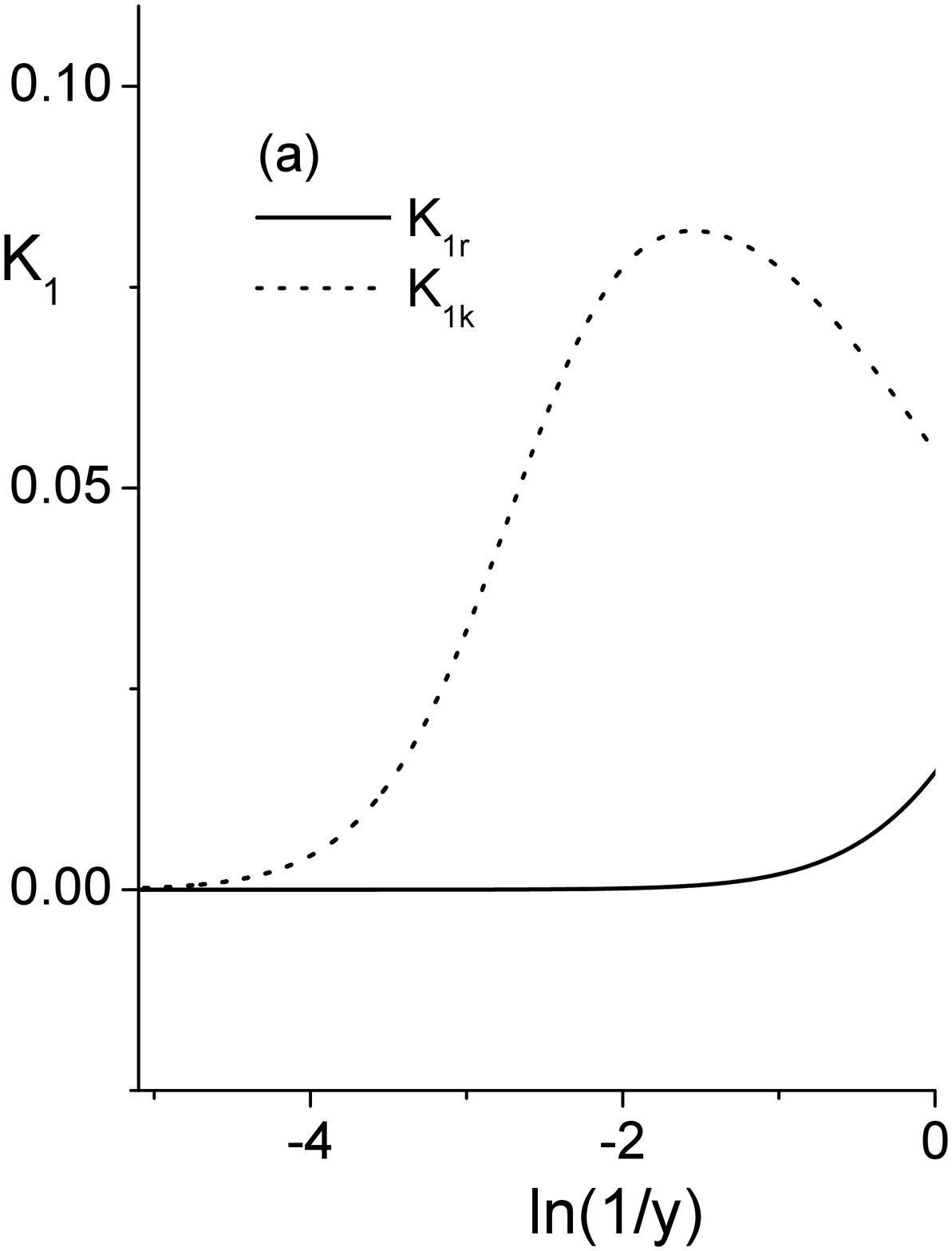}
 \hspace{0.3cm}
 \includegraphics[height=5.0cm,width=3.5cm]{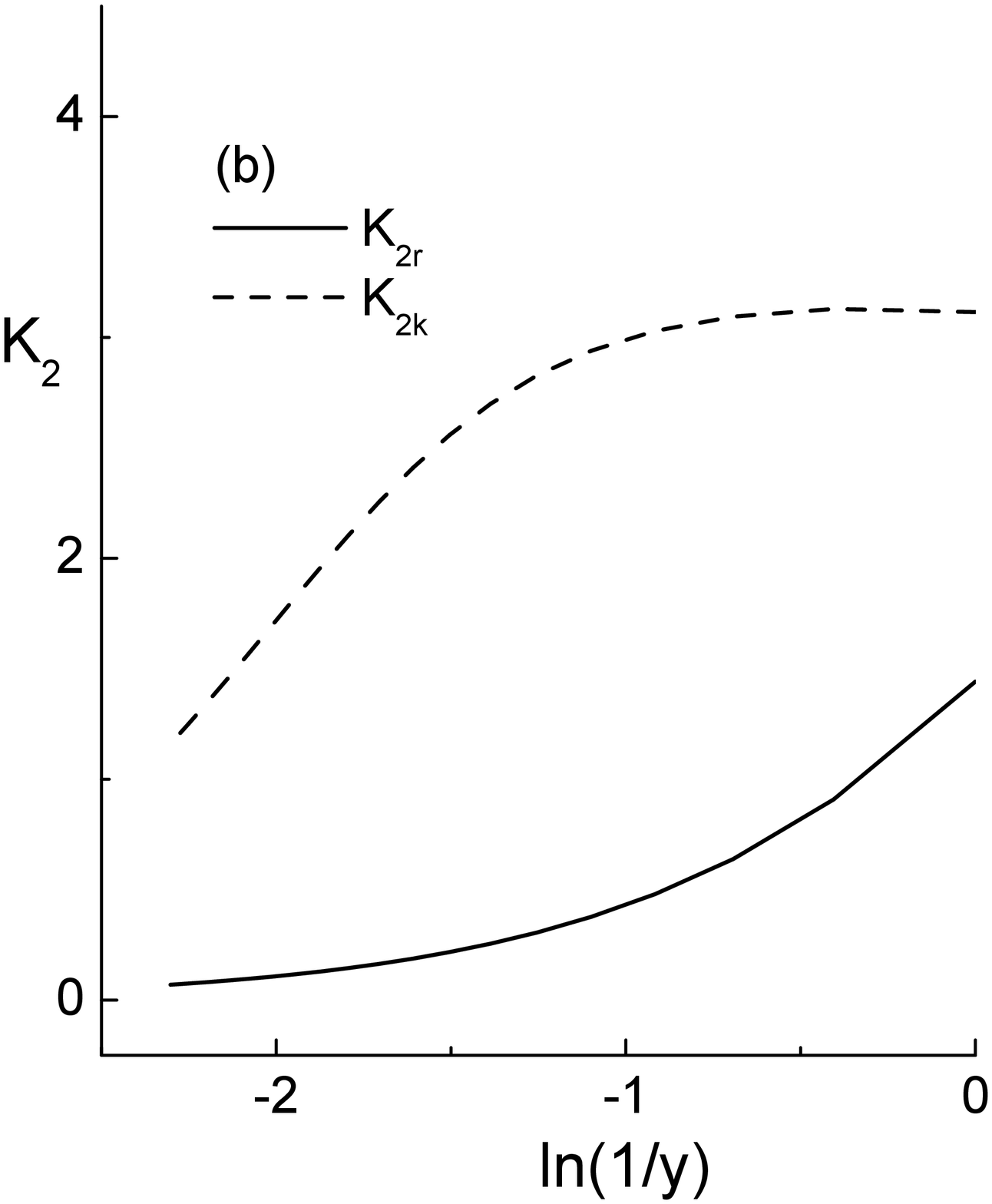}
 \caption{(a) One- body Kullback-Leibler relative entropy both in
coordinate- and momentum-space (b) Two-body Kullback-Leibler
relative entropy both in coordinate- and momentum-space, in a
trapped Bose gas.}\label{fig:fig10}
\end{figure}

A well-known concept in information theory is the distance between
the probability distributions $\rho_i^{(1)}$ and $\rho^{(2)}$, in
our case the correlated and the uncorrelated distributions
respectively. A measure of distance is the Kullback-Leibler
relative entropy $K$ defined previously. The correlated and
uncorrelated cases are compared for the one-body case $(K_1)$  and
for the two-body case $(K_2)$ in Fig. \ref{fig:fig10}, decomposing
in position- and momentum-spaces according to
(\ref{eq:equ17})-(\ref{eq:equ20}). It is seen that $K_{1r}$,
$K_{2r}$ increase as the strength of SRC increases, while
$K_{1k}$, $K_{2k}$ have a maximum at a certain value of
$\ln{(\frac{1}{y})}$ depending on the system under consideration.

\begin{figure}[h]
 \centering
 \includegraphics[height=5.0cm,width=3.5cm]{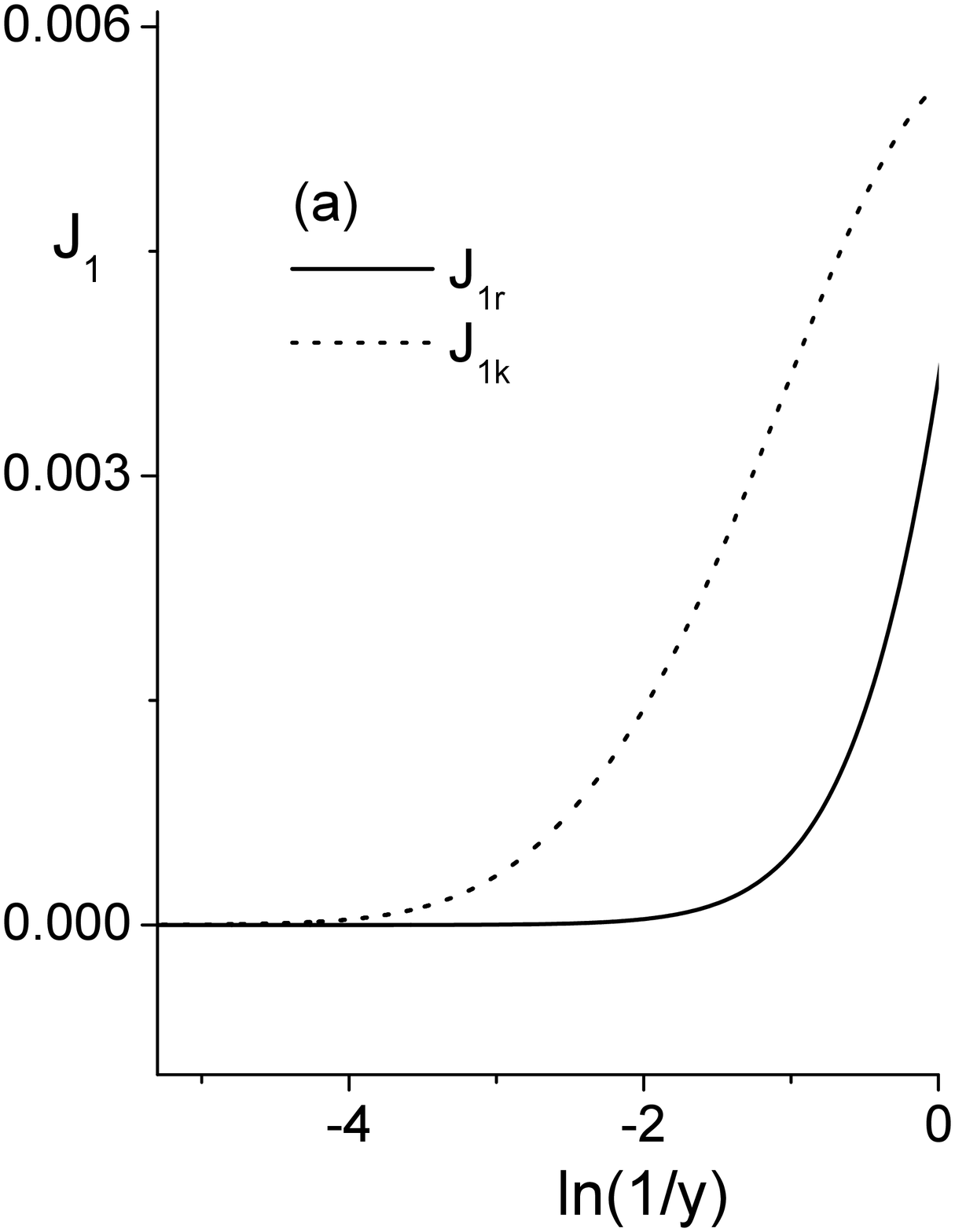}
 \hspace{0.3cm}
 \includegraphics[height=5.0cm,width=3.5cm]{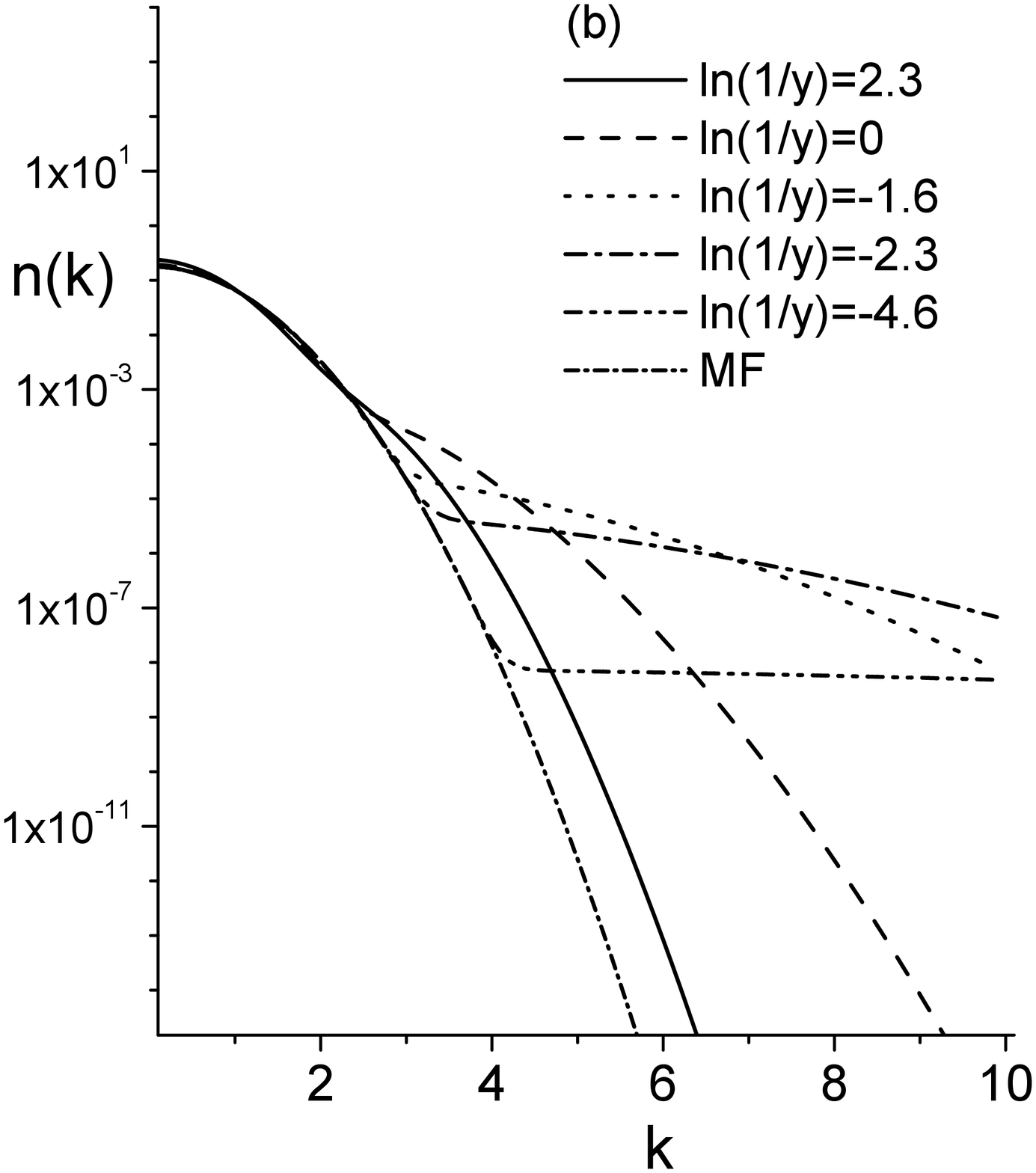}
 \caption{ (a) One-body Jensen-Shannon divergence entropy both in coordinate-
and momentum-space, in a trapped Bose gas (b) Momentum
distribution $n(k)$ of $^4$He for various values
 of the correlation parameter $\ln{(\frac{1}{y})}$. The case MF (mean field)
 corresponds to the uncorrelated case $(y\rightarrow \infty).$ }\label{fig:fig11}
\end{figure}

Calculations are also carried out for the Jensen-Shannon
divergence for one-body density distribution ($J_1$ entropy) as
function of $\ln{(\frac{1}{y})}$, decomposed in position- and
momentum- spaces (Fig. \ref{fig:fig11}). We observe again that
$J_1$ increases with the strength of SRC in position-space, while
in momentum-space there is a maximum for a certain value of
$\ln{(\frac{1}{y})}$. It is verified that $0<J<\ln{2}$ as expected
theoretically \cite{Majtey05}.

It is noted that the dependence of the various kinds of
information entropy on the correlation parameter
$\ln{(\frac{1}{y})}$ is studied up to the value
$\ln{(\frac{1}{y})}=0$ $(y=1)$, which is already unrealistic
corresponding to strong SRC. In addition, lowest order
approximation does not work well beyond that value. In this case
three-body terms should be included but this prospect is out of
the scope of the present work.

For very strong SRC the momentum distribution $n(k)$ exhibits a
similar behavior with the mean field $(y\rightarrow \infty)$. This
is illustrated in Fig. \ref{fig:fig5}, where we present $n(k)$ for
various values of $\ln{(\frac{1}{y})}$. It is seen that for small
and large SRC the tail of $n(k)$ disappears. That is why for small
and large SRC the relative entropy ($K_{1k}$ and $J_{1k}$) is
small, while in between shows a maximum (Fig. \ref{fig:fig10}). A
similar trend of $n(\textbf{k}_1,\textbf{k}_2)$ for large SRC
explains also the maximum of the relative entropy $K_{2k}$.

\subsection{Conclusions and comments}\label{sub:sec4-7}

Our main conclusions are the following \cite{MoustaChatz05}
\begin{itemize}
\item[(i)] Increasing the SRC (i.e. the parameter
$\ln{(\frac{1}{y})}$) the information entropies $S$, $O$, $K$ and
$J$ increase. A comparison leads to the conclusion that the
correlated Bose gas has larger values of entropies than the
uncorrelated one.

\item[(ii)] There is a relation of $S_r$ with $\sqrt{\langle r^2
\rangle}$ and $S_k$ with $\langle T \rangle$  in the sense that
they have the same behavior as a function of the correlation
parameter $\ln{(\frac{1}{y})}$. These results can lead us to
relate the theoretical quantities $S_r$ and $S_k$ with
experimental ones like charge density distribution, and momentum
distribution, radii, etc. A recent paper addresses this problem
\cite{Psonis05}.

\item[(iii)] The relations $S_2=2 S_1$ and $O_2=O_1^2$ hold
exactly for the uncorrelated densities in trapped Bose gas while
the above relations are almost exact in the case of correlated
densities. In previous work we proposed the universal relation
$S_1=S_r+S_k=a+b\,\ln{N}$ where $N$ is the number of particles of
the system either fermionic (nucleus, atom, atomic cluster) or
bosonic (correlated atoms in a trap). Thereby, in the general case
(including correlations among the atoms)
\[
  S_2\simeq 2 S_1=2(a+b\,\ln{N}).
\]

For the generalized $N$-body uncorrelated distributions
$\rho(\textbf{r}_1,\textbf{r}_2,\ldots,\textbf{r}_N)$ and
$n(\textbf{k}_1,\textbf{k}_2,\ldots,\textbf{k}_N)$ the relation
\[
  S_N=N S_1=N\,(a+b\,\ln{N}).
\]
holds exactly.

It is conjectured that it holds approximately for correlated
systems (which has still to be proved for $N\geq 3$).

\item[(iv)] The entropic uncertainty relation (EUR) is
\[
  S=S_r+S_k\geq 6.434.
\]
It is well-known that the lower bound is attained for a Gaussian
distribution (i.e. the case of uncorrelated Bose gas). In all
cases studied in the present work EUR is verified.

A final comment seems appropriate. In general, the calculation of
$\rho({\bf r}_1,{\bf r}_2)$ and $n({\bf k}_1,{\bf k}_2)$ is a
problem very hard to be solved in the framework of short range
correlations. In the present work we tried to treat the problem in
an approximate but self-consistent way in the sense that the
calculations of $\rho({\bf r}_1,{\bf r}_2)$ and $n({\bf k}_1,{\bf
k}_2)$ are based in the same $\rho({\bf r}_1,{\bf r}_2;{\bf
r}_1',{\bf r}_2')$, which is the generating function of the above
quantities. As a consequence the information entropy
$S_2=S_{2r}+S_{2k}$ is derived also in a self-consistent way and
there is a direct link between $S_{2r}$ and $S_{2k}$, as well as
the other kinds of information entropies which are studied in the
present work.
\end{itemize}

\section{Quantum-Information properties based on Gross-Pitaevskii
equation}\label{sec:sec5}

The ground-state properties of the condensate, for weakly
interacting atoms, are explained quite successfully by the
non-linear equation, known as Gross-Pitaevskii  (GP) equation, of
the form
\begin{equation}
\left[-\frac{\hbar^2}{2m} \nabla^2 + \frac{1}{2}m\omega^2 r^2+
N\frac{4\pi \hbar^2 a_0 }{m} | \psi({\bf r}) | ^2\right] \psi({\bf
r})=\mu \,\psi({\bf r}), \label{gros-pit}
\end{equation}
where $N$ is the number of the atoms, $m$ is the atomic mass,
$a_0$ is the scattering length of the interaction and $\mu$ is the
chemical potential \cite{Dalfavo99}. This equation has the form of
a non-linear stationary Schr\"odinger equation, and it has been
solved for several types of traps using various numerical methods
\cite{Dalfavo96,Edwards95,Cerboneschi}. The presence of the third
term, which is linear in $N$ is responsible for the dependence of
the gas parameter $\chi$  on the density of the system.

 Eq. (\ref{gros-pit}) was solved numerically
in Ref. \cite{Massen02} for trapped boson-alkali atoms in two
cases. For a system of ${}^{87}$Rb atoms with parameter
$b_0=12180$ \AA $\,$ (angular frequency $\omega_0/2\pi=77.78$ Hz)
and scattering length $a_0=52.9$ \AA$\,$ \cite{Fabrocini99}, and
for a system of $^{133}$Cs atoms with parameters $b_0=27560$ \AA
$\,$ ($\omega_0/2\pi=10$ Hz) and $a_0=32$ \AA$\,$ \cite{Schuck00}.
 In these cases the effective
atomic size is small compared both to the trap size and to the
interatomic distance ensuring the diluteness of the gas.

There is in atomic physics a connection of $S_{r}$ and $S_{k}$
with the total kinetic energy $T$ and mean square radius of the
system through rigorous inequalities derived using the EUR
\cite{Gadre87,Gadre85}
\begin{eqnarray}
 \label{Sr-ineq}
 &&S_r({\rm min}) \le S_r \le S_r({\rm max})\\
 \label{Sk-ineq}
 &&S_k({\rm min}) \le S_k \le S_k({\rm max})\\
 \label{S-ineq}
 &&S({\rm min}) \le S \le S({\rm max}).
\end{eqnarray}

The lower and upper limits are written here more conveniently in
the following form, for density distributions normalized to one:
\begin{eqnarray}
S_r({\rm min})&=& \frac{3}{2}(1+\ln \pi)
           -\frac{3}{2} \ln \left( \frac{4}{3} T \right) \nonumber\\
S_r({\rm max})&=& \frac{3}{2}(1+\ln \pi)
           +\frac{3}{2}\ln \left(\frac{2}{3}\langle r^2 \rangle
           \right),
 \label{Sr-min}
\end{eqnarray}
\begin{eqnarray}
S_k({\rm min})&=& \frac{3}{2}(1+\ln \pi)
           -\frac{3}{2}\ln \left( \frac{2}{3}\langle r ^2\rangle \right) \nonumber\\
S_k({\rm max})&=& \frac{3}{2}(1+\ln \pi)
           +\frac{3}{2}\ln \left(\frac{4}{3} T \right),
            \label{Sk-min}
\end{eqnarray}
\begin{eqnarray}
S({\rm min})&=& 3(1+\ln \pi)  \nonumber\\
S({\rm max})&=& 3(1+\ln \pi)
        + \frac{3}{2}\ln \left(\frac{8}{9}\langle r^2\rangle T
        \right).
 \label{S-min}
\end{eqnarray}

In Ref. \cite{Massen01} it was  verified numerically that the
above inequalities hold for nuclear density distributions and
valence electron distributions in atomic clusters. We also found a
link of $S$ with the total kinetic energy of the system $T$, and a
relationship of Shannon's information entropy in position-space
$S_{r}$ with an experimental quantity i.e. the rms radius of
nuclei and clusters.

It has been verified numerically \cite{Massen02} that inequalities
(\ref{Sr-ineq}), (\ref{Sk-ineq}) and (\ref{S-ineq}) hold for
correlated bosonic systems as well, i.e. trapped boson-alkali
atoms $^{87}$Rb and $^{133}$Cs. That is shown in Table
\ref{tbl:table2} for $^{87}$Rb bosonic system. An analogous table
can be displayed for $^{133}$Cs. It is noted that for large $N$,
$S_k$ may become negative, but the important quantity is the net
information content $S=S_r+S_k$ of the system which is positive.
We employed density distributions $\rho({\bf r})$ and $n({\bf k})$
for bosons derived by solving numerically the GP equation
(\ref{gros-pit}). It is also seen that the right-hand-side of
inequality (\ref{Sr-ineq}) is nearly an equality. Thus there
exists a relation between $S_r$ and $ T $ for bosons as well as
for nuclei \cite{Massen01}.


\begin{table}[pt]
\centering
\begin{tabular}{@{}r c c c c c c c c c@{}}
\hline    $N$ & $S_r$& $S_r\,\,\,$& $S_r$& $S_k$& $S_k\,\,\,$
& $S_k$& $S$ &$S\,\,\,\,$& $S$ \\
\hline
5$\times 10^2$ & 3.797 &3.834 &3.845 &2.590 &2.630 &2.637 &6.434 &6.465 &  6.482\\
$10^3$ &4.027 &4.100 &4.120 & 2.314 & 2.394 & 2.408 &6.434 &6.494 &6.528\\
3$\times 10^3$ &4.437 &4.599 &4.640 & 1.794 & 1.963 & 1.997 &6.434 &6.562 &6.637 \\
5$\times 10^3$ &4.641 &4.855 &4.907 & 1.527 & 1.746 & 1.794 &6.434 &6.601 &6.701 \\
7$\times 10^3$ &4.778 &5.029 &5.090 & 1.345 & 1.598 & 1.657 &6.434 &6.627 &6.746 \\
$10^4$ &4.925 &5.219 &5.287 & 1.148 & 1.437 & 1.509 &6.434 &6.655 &6.796 \\
5$\times 10^4$ &5.615 &6.113 &6.211 & 0.223 & 0.667 & 0.819 &6.434 &6.780 &7.030 \\
$10^5$ &5.922 &6.511 &6.619 &-0.185 & 0.317 & 0.512 &6.434 &6.828 &7.132\\
5$\times 10^5$ & 6.654 &7.452 &7.577 &-1.142 &-0.533 &-0.220 &6.434 &6.919 &7.357 \\
$10^6$ &6.993 &7.864 &7.992 &-1.557 &-0.920 &-0.560 &6.434 &6.943 &7.432 \\
\hline
\end{tabular}
\caption{Values of $S_r$, $S_k$ and $S$ versus the number of
particles $N$ for ${}^{87}$Rb bosonic system (see inequalities
(\ref{Sr-ineq}),
 (\ref{Sk-ineq}), and (\ref{S-ineq})) \cite{Massen02}.}\label{tbl:table2}
\end{table}

\begin{figure}[t]
\centering
\includegraphics[height=5.0cm,width=4.cm]{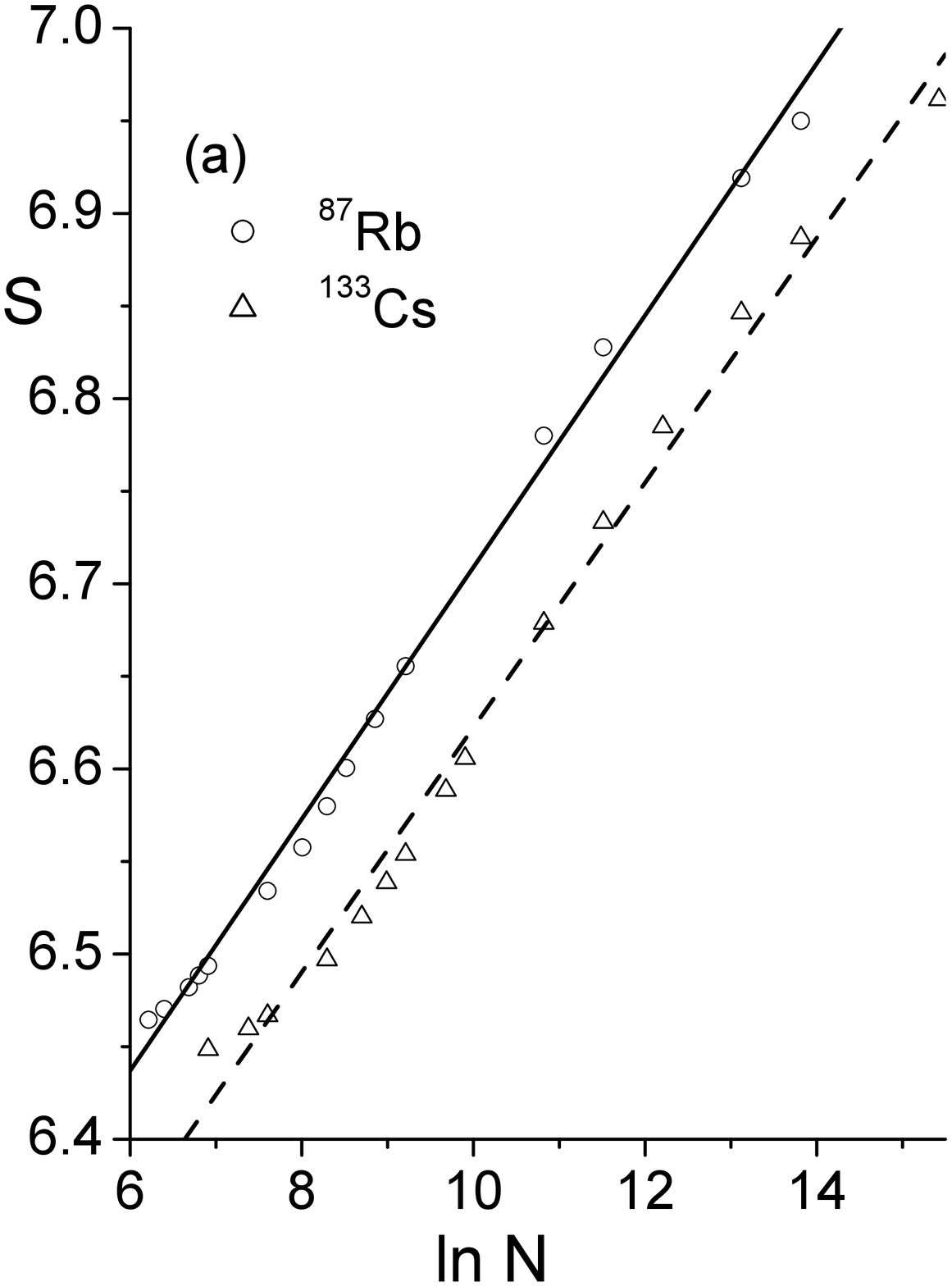}
\hspace{0.5cm}
\includegraphics[height=5.0cm,width=4.cm]{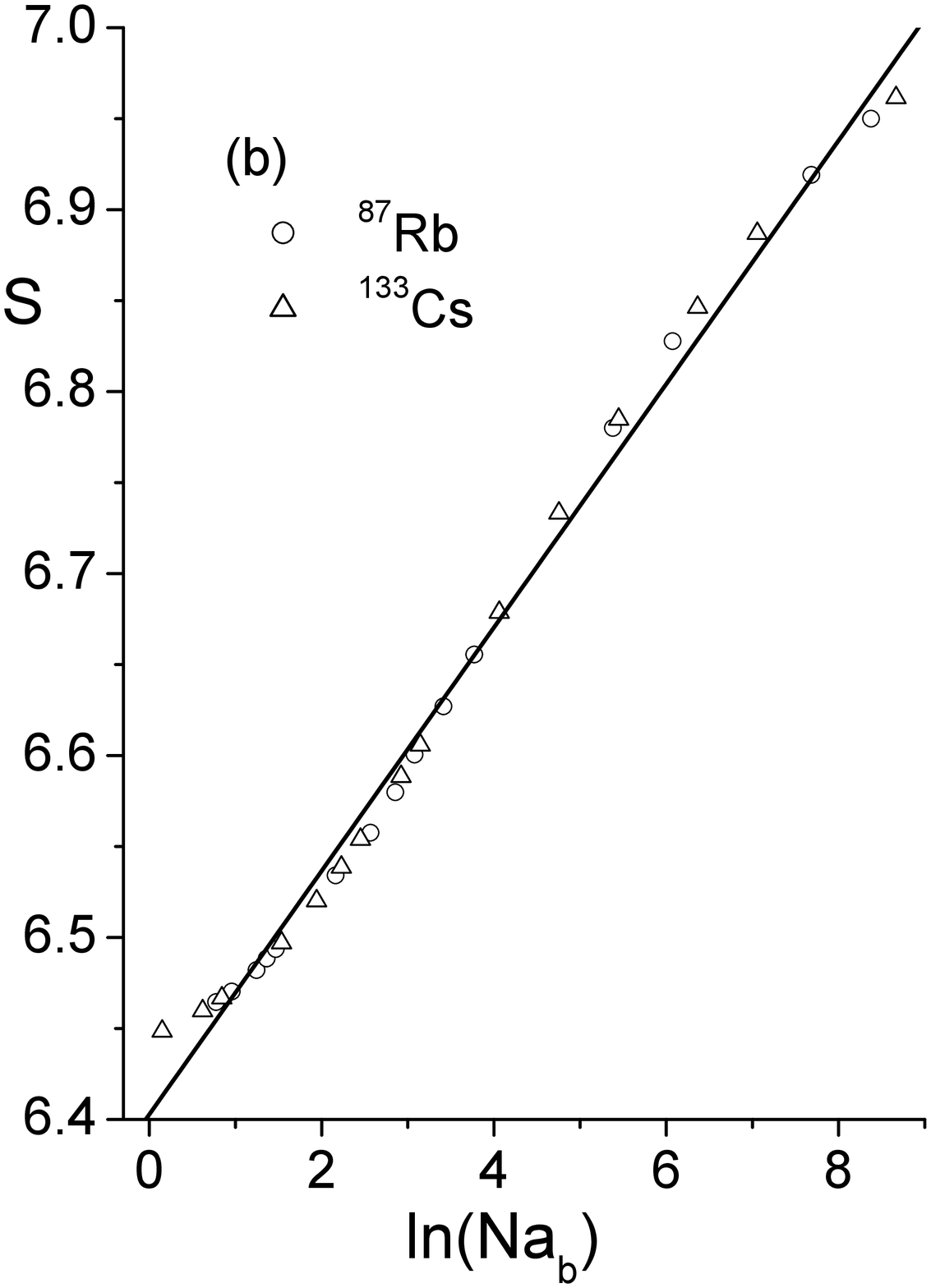}
\caption{The information entropy $S$ (a) versus $\ln N$ and (b)
versus $\ln (Na_b)$ for the bosonic systems according to the
fitted expressions (\ref{S-ln1}) and (\ref{SrSk-3}) respectively.
The open circles ($^{87}$Rb) and triangles ($^{133}$Cs) come from
the numerical solution of the GP equation.} \label{fig:fig12}
\end{figure}

In Ref. \cite{Massen02} we addressed the problem of finding
$S_{r}$ and $S_{k}$ (i.e. the extent of $\rho({\bf r})$ and
$n({\bf k})$) for bosonic many-body systems in order to compare
with corresponding results for fermionic systems. First we review
the results of Ref. \cite{Massen98} for systems of fermions where
we proposed a universal property for $S$ for the density
distributions of nucleons in nuclei, electrons in atoms and
valence electrons in atomic clusters. This property has the form
\begin{equation}
S=a+b \ln{N},
 \label{S-ln1}
\end{equation}
where the parameters $a$ and $b$ depend on the system under
consideration. The values of the parameters are the following
\begin{eqnarray}
a=5.325,& \quad b=0.858 & \quad ({\rm nuclei}) \nonumber\\
a=5.891,& \quad b=0.849 & \quad ({\rm atomic \ clusters})\\
a=6.257,& \quad b=1.007 & \quad ({\rm atoms}).  \nonumber
\label{eq-8}
\end{eqnarray}
Next \cite{Massen02}, we verified (\ref{S-ln1}) employing
densities $\rho (r)$ and $n(k)$ for trapped bosons solving the GP
equation (\ref{gros-pit}). $\rho (r)$, derived in this way, and
$n(k)$ derived by Fourier transform of $\psi(r)$, were inserted
into equations (\ref{eq:equ3}) and (\ref{eq:equ4}) to find the
values of $S_r$, $S_{k}$ and $S=S_{r}+S_{k}$ as functions of the
number of bosons $N$. The results are shown in Fig.
\ref{fig:fig12}(a). The circles and the triangles correspond to
the calculated values for the bosonic systems ${}^{87}$Rb and
$^{133}$Cs, respectively, while the lines to the fitted form of
Eq. (\ref{S-ln1}) where
\begin{eqnarray}
a=6.0291, & \quad b=0.0678, & \quad 5\times 10^2 < N < 10^6 \,\,\, (^{87}{\rm Rb})\nonumber\\
a=5.9614,  & \quad b=0.0661,  & \quad 10^3 < N <5\times 10^6
\,\,\,   (^{133}{\rm Cs}). \label{ab-SrSk-2}
\end{eqnarray}

From the values of the parameters $a$ and $b$, for the two systems
we examined, and from the strength of the interatomic interaction,
which is proportional to $N a_0/m$ we can conclude that for the
same trap (same value of $b_0$) and for different values of the
scattering length or/and  of the atomic mass, the linear
dependence of $S$ on $\ln N$ does not change. The only change will
be in the values of the parameter $a$ of Eq. (\ref{S-ln1}). Thus
the change of the scattering length or/and of the atomic mass will
produce a parallel displacement of the lines of Fig.
\ref{fig:fig12}(a). From the values of the parameter $b$ of Eq.
(\ref{S-ln1}) which give the slope of the lines of Fig.
\ref{fig:fig12}(a) corresponding to different sizes of the trap
$b_0$ and because the GP equation can be written in the form
\begin{equation}
\left[-\frac{1}{2} \nabla_{r_b}^2 + \frac{1}{2} r_b^2+ 4\pi N a_b
| \psi_b({\bf r}_b) | ^2\right] \psi_b({\bf r}_b)=\mu_{b}
\psi_b({\bf r}_b), \label{gros-pit2}
\end{equation}
where $r_b=r/{b_0}$, $a_b=a_0/{b_0}$,  $\mu_{b}=\mu /
{\hbar\omega_0}$ and $\psi_b({\bf r}_b)=b_0^{3/2} \psi({\bf r})$,
we should conclude that there is a parallel displacement of the
lines for the various values of $b_0$. That leads us to fit the
numerical values of $S$, for the two systems we examined, using
the formula
\begin{equation}
S= a+b \ln\left( N a_b \right). \label{SrSk-3}
\end{equation}
The new values of $a$ and $b$ are now $a=6.3976$ and  $b=0.0678$
for ${}^{87}$Rb and $a=6.4081$ and $b=0.0661$ for ${}^{133}$Cs.
As the two lines are almost the same we use the same parameters
$a$ and $b$ for the two bosonic systems
\begin{equation}
a=6.4028,\quad b=0.0669 \label{ab-SrSk-3b}
\end{equation}
which are the mean values of the corresponding parameters of the
two systems. The results are shown in Fig. \ref{fig:fig12}(b). It
is seen that the numerical values of $S$ for the two bosonic
systems are very close to those calculated from Eq. (\ref{SrSk-3})
with the parameters $a$ and $b$ given by Eq. (\ref{ab-SrSk-3b}).

\begin{figure}[h]
\centering
\includegraphics[height=5.0cm,width=4.cm]{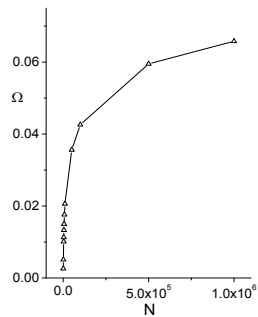}
 \caption{The order parameter $\Omega$ as a function
of the number of particles $N$ for ${}^{87}$Rb bosonic
system.}\label{fig:fig13}
\end{figure}

The results of $S$ and $S({\rm max})$, displayed in Table
\ref{tbl:table2}, allowed us to calculate the order parameter
$\Omega$ (relation (\ref{omega})) as function of the number of
particles $N$ in a system of trapped correlated boson-alkali
atoms. The dependence of $\Omega$ on $N$ is shown  in Fig.
\ref{fig:fig13} for ${}^{87}$Rb (an analogous figure can be
displayed for ${}^{133}$Cs). It is seen that $\Omega$ is an
increasing function of $N$. A similar trend has been observed in
Fig. 1 of Ref. \cite{Panos01b}, where $\Omega(N)$ was calculated
for nucleons in nuclei and valence electrons in atomic clusters.
As stated in \cite{Panos01b}, our result is in a way
counter-intuitive and indicates that as particles are added in a
correlated quantum-mechanical system, the system becomes more
ordered. The authors in \cite{Landsberg98} studied disorder and
complexity in an ideal Fermi gas of electrons. They observed that
for a small number of electrons the order parameter $\Omega$ is
small, while $\Omega$ increases as one pumps electrons into the
system and the energy levels fill up.

\subsection{Shannon's Information Entropy Using Density Distribution
of Correlated Bosons, Results and Discussion}\label{sub:sub5-1}

\begin{figure}[h]
\centering
\includegraphics[height=5.0cm,width=4.0cm]{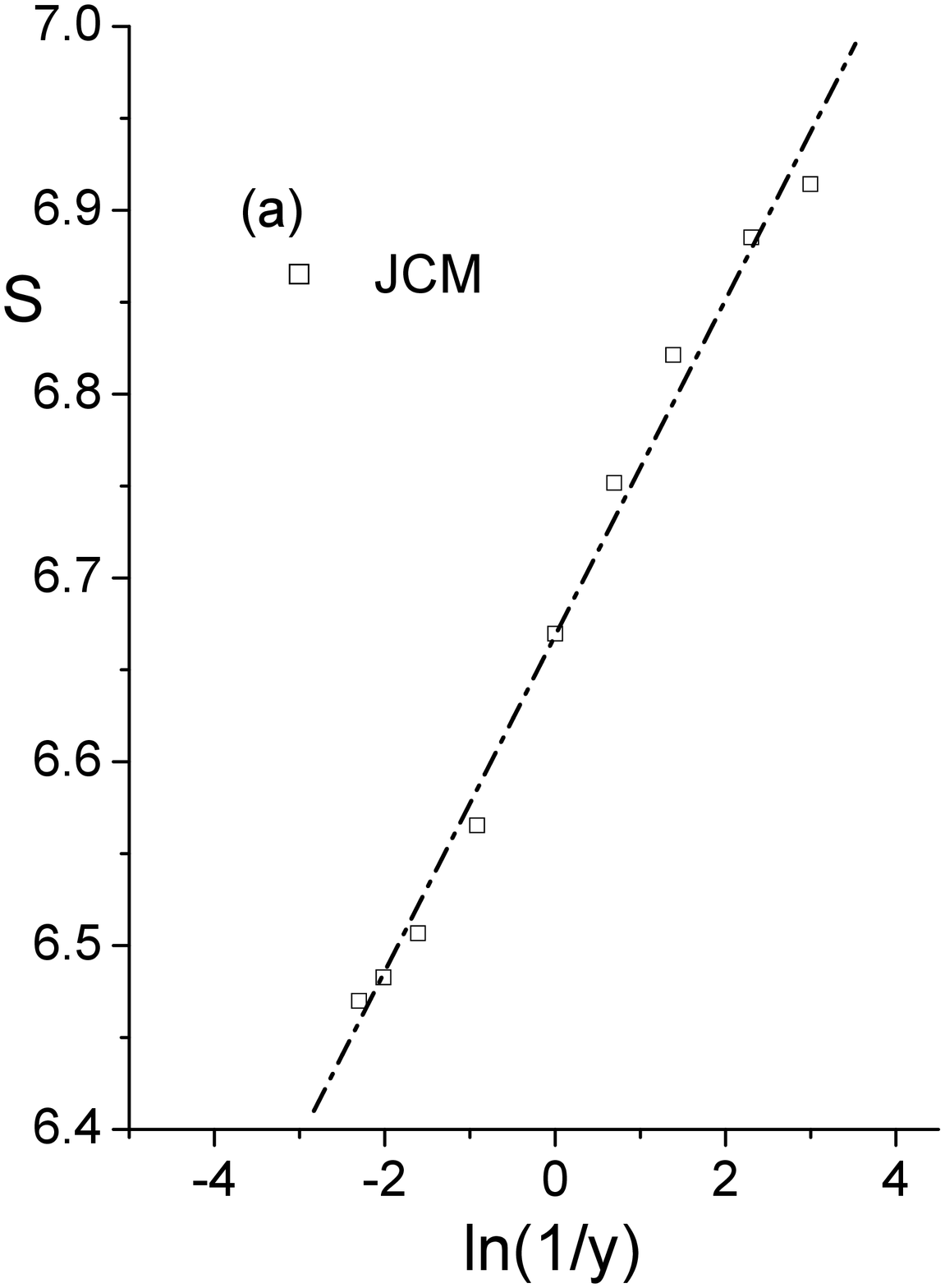}
\hspace{0.5cm}
\includegraphics[height=5.0cm,width=4.cm]{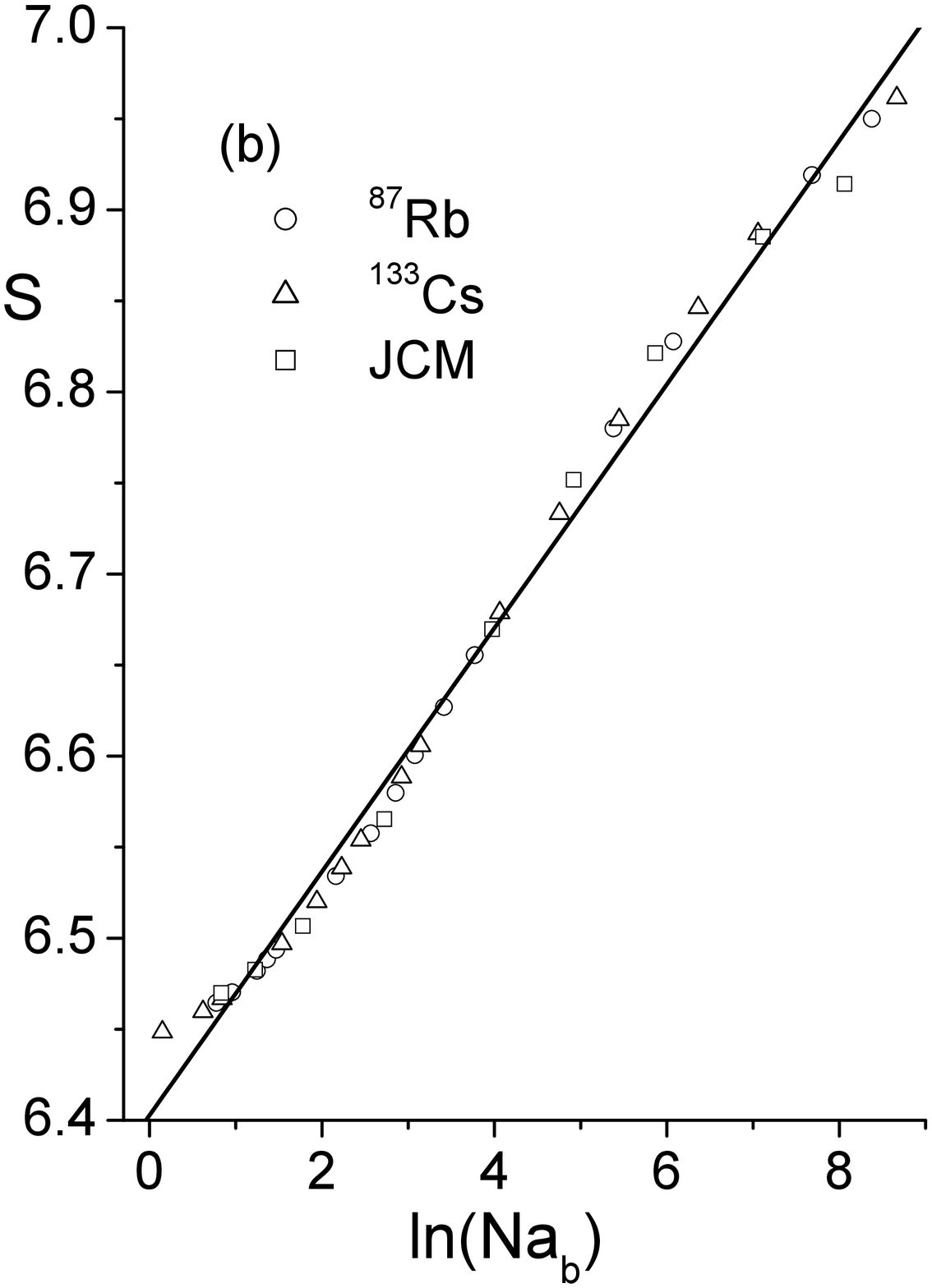}
\caption{(a) The information entropy $S$ versus the correlation
parameter $y$ for a bosonic system. The line corresponds to the
fitted expression (\ref{SrSk-4}), while the open squares to values
found using the JCM. (b) The information entropy $S$ versus $\ln
(Na_b)$ for the bosonic systems according to the fitted expression
(\ref{SrSk-3}). The open circles (${}^{87}$Rb) and triangles
(${}^{133}$Cs) come from the numerical solution of the GP
equation, while the open squares correspond to the
JCM.}\label{fig:fig14}
\end{figure}

From the analytical expressions of $\rho(r)$ and $n(k)$  (Eqs.
(\ref{cluster-nr}) and (\ref{cluster-nk}) respectively) for a
correlated bosonic system, the information entropy $S$ can be
found using Eqs. (\ref{eq:equ3}) and (\ref{eq:equ4}). The values
of $S$ as function of the correlation parameter $\frac{1}{y}$ are
shown in Fig. \ref{fig:fig14}(a). The open squares correspond to
the calculated values of $S$ while the line to the fitting
expression
\begin{equation}
 S=a+b\ln(1/y),
 \label{SrSk-4}
 \end{equation}
 where
 \begin{eqnarray}
 a=6.6687,& \quad b=0.0913,&\,\,\, 0.05 \le y \le 10.
  \label{ab-srsk-4}
 \end{eqnarray}

We have fitted our numerical results for $0.05 \leq y \leq 10$. As
mentioned before large values of $y$ ($y > 10$) correspond to a
Gaussian distribution, while for $y\lesssim 0.05$ (very strong
correlations) higher order terms must be included in the expansion
of the density.

It should be noted that in the present approach there is not a
direct dependence between the condensation and the number of the
atoms. The inter-particle correlations are incorporated in the
mean field only by the correlation function which, in some way,
depends on the effective size of the atoms. We could find that
dependence making the assumption that the correlated parameter $y$
depends on $N$ and $a_b$ through the relation
\begin{equation}
\frac{1}{y}= \left(\lambda_1 Na_b \right)^{\lambda_2}
\label{inv-y}
\end{equation}
 and try to find $\lambda_1$ and $\lambda_2$ equating the rhs of
 Eqs. (\ref{SrSk-3}) and (\ref{SrSk-4}). In this way $\lambda_1$
 and  $\lambda_2$ can be found as functions of the parameters $a$
 and $b$ of Eqs. (\ref{SrSk-3}) and (\ref{SrSk-4}) having the
 forms
 \begin{equation}
 \lambda_1 ={\rm e}^{(a_1-a_2)/b_1},\quad
 \lambda_2 =b_1/b_2
 \label{a12-b12}
 \end{equation}
where $a_1$ and $b_1$ are the values of the parameters $a$ and $b$
of Eq. (\ref{ab-SrSk-3b})
 and $a_2$ and $b_2$ the parameters of Eq. (\ref{ab-srsk-4}).
 The numerical values of
$\lambda_1$ and $\lambda_2$ are: $\lambda_1=0.0188$ and
$\lambda_2=0.7330$.

The values of $S$ for the bosonic systems ${}^{87}$Rb and
$^{133}$Cs found in Jastrow correlation method (JCM), versus $\ln
(Na_b)$ (calculated from Eqs. (\ref{inv-y}) and (\ref{a12-b12})),
are shown in Fig. \ref{fig:fig14}(b) with open squares. It is seen
that the two bosonic systems studied with the GP theory and the
bosonic system studied with the JCM give very similar results for
$S$. It seems that the information entropy $S$ for the bosonic
systems depends only on $\ln(Na_b)$.

\section{Summary}\label{sec:sec6}
The effect of the interparticle correlations between Bose atoms at
zero temperature is examined using a phenomenological way to
incorporate the atomic correlations. This is made by introducing
the Jastrow correlation function in the two-body density matrix.
Analytical expressions are found for the one- and the two-body
density and momentum distribution, mean-square radius, kinetic
energy and static structure factor. The introduction of
correlations changes the shape of the density and momentum
distributions compared with the Gaussian one, corresponding to the
harmonic oscillator model. There is a decrease of the density
distribution in the central region of the atomic system while the
momentum distribution increases in the region of small $k$ and
thus there is a decrease of the mean kinetic energy of the system.
In addition the natural orbitals and the natural occupation
numbers have been calculated and consequently the condensate
fraction has been obtained for different values of the parameter
$y$. A theoretical calculation of the static structure factor is
reported also by applying two trial forms for the radial
distribution function. Our results are compared with recent
experimental data concerning trapped Bose gas. By applying
suitable parametrization the experimental data are reproduced
quite well.

Various kinds of quantum information properties of the trapped
Bose gas are calculated i.e. the Shannon and Onicescu information
measures for the correlated and uncorrelated cases which are
compared as functions of the strength of the short range
correlations. It can be seen that increasing the short range
correlations the information entropies $S$ and $O$ increase. There
is a relation between $\sqrt{\langle r^2 \rangle}$ and $S_r$ and
between $\langle T \rangle$ and $S_k$. It is  also conjectured
that the relation $S_N=N(a+b\,\ln{N})$ holds approximately for the
correlated system. The Gross-Pitavskii equation is solved in order
to calculate the information properties of the Bose gas from
another point of view. It is concluded that the Shannon
information entropy obeys the functional form $S=a+b\,\ln{N}$.
Finally it is shown that Landsberg's order parameter $\Omega$ is
an increasing function of the number of Bose atoms $N$.

\section*{Acknowledgments}

The work of K.~Ch.~Chatzisavvas and C.~P.~Panos was supported by
Herakleitos Research Scholarships (21866) of
$\textrm{E}\Pi\textrm{EAEK}$ and the European Union while the work
of S.~E.~Massen, Ch.~C.~Moustakidis and C.~P.~Panos  was supported
by the Pythagoras II Research project (80861) of
$\textrm{E}\Pi\textrm{EAEK}$ and the European Union.

\end{document}